\documentstyle[psfig,preprint,graphicx,tighten,eqsecnum,floats,aps]{revtex}

\makeatletter
\@addtoreset{equation}{section}

\makeatother
\newcommand{\wwedge}[0]{\rlap{$\wedge$}\!\!\wedge}
\newcommand{\dd}[0]{\rlap{$\mathrm{d}$}\hspace{0.3ex}\mathrm{d}}
\newcommand{\grad}[0]{\nabla\!}

\newcommand{\hateq}[0]{\mathrel{\widehat\mathalpha{=}}}
\newcommand{\D}{{\mathcal{D}}}

\newcommand{\Lie}[0]{{\cal L}\, }

 \makeatletter
 \newcommand{\pback}[1]{{
   \let\@rrow=\leftarrowfill
   \mathchoice{\AIN@stemPullBack{#1}{\@rrow}}{\AIN@stemPullBack{#1}{\@rrow}}
     {\AIN@indxPullBack{#1}{\@rrow}}{\AIN@indxPullBack{#1}{\@rrow}}}
   \vphantom{#1}}

 \newcommand{\AIN@stemPullBack}[2]{
   \vtop{\mathsurround=0pt
   \ialign{##\crcr$\textstyle{#1}\strut$\crcr
     \noalign{\kern-0.4ex\nointerlineskip}{\tiny#2}\crcr}}}

 \newcommand{\AIN@indxPullBack}[2]{
   \vtop{\mathsurround=0pt
   \ialign{##\crcr\hfil$\scriptstyle{#1}$\hfil\crcr
     \noalign{\kern+0.4ex\nointerlineskip}{\tiny#2}\crcr}}}

\makeatother

\newcommand{\into}[0]{\cdot}
\newcommand{\Tr}[0]{\mathrm{Tr}}

\newcommand{\RePt}[1]{{\mathrm{Re}} \left[ #1 \right]}
\newcommand{\ImPt}[1]{{\mathrm{Im}} \left[ #1 \right]}

\newcommand{\twoeps}[0]{{}^2\!\epsilon}

\newcommand{\emA}{{\mathbf{A}}}
\newcommand{\emF}{{\mathbf{F}}}

\newcommand{\emT}{{\mathbf{T}}}

\newcommand{\ymA}{{\mathbf{A}}}
\newcommand{\ymF}{{\mathbf{F}}}
\newcommand{\ymT}{{\mathbf{T}}}
\newcommand{\modF}{\!\mid\!\! \ymF \!\!\mid\!}
\newcommand{\modstarF}{\mid \!\dual\ymF \!\mid}
\newcommand{\ymphi}{\Phi^{\YM}}
\newcommand{\YM}{{\mathrm{YM}}}
\newcommand{\ymD}{{\mathbf{D}}}

\newcommand{\I}{\mfs{I}}

\newcommand{\man}{{\mathcal{M}}}

\newcommand{\ADM}{{\mathrm{ADM}}}
\newcommand{\dual}{{}^\star}
\def\S{\mathcal{S}}

\newcommand{\half}[0]{\frac{1}{2}}

\let\puto=\overcirc
\def\kt{{\kappa}_{(t)}\,}

\def\l{\ell}
\def\ls{{(\l)}}

\def\Pt{\Phi_{(t)}\,}

\def\bar{\overline}
\def\ba{\begin{eqnarray}}
\def\ea{\end{eqnarray}}
\def\be{\begin{equation}}
\def\ee{\end{equation}}
\def\={\hateq}
\def\puto#1{\rlap{\raise.5ex\hbox{\char'27}}{#1}}
\def\I{\cal I}
\def\n{{\vec n}}

\newcommand  {\Rbar} {{\mbox{\rm$\mbox{I}\!\mbox{R}$}}}

\begin{document}
\title{Isolated Horizons: Hamiltonian Evolution \\
    and the First Law}
\author {Abhay Ashtekar $^{1,2}$ \thanks{E-mail:
ashtekar@gravity.phys.psu.edu}, Stephen Fairhurst $^{1}$
\thanks{E-mail: fairhurs@phys.psu.edu} and Badri Krishnan $^{1}$
\thanks{E-mail: krishnan@phys.psu.edu}}

\address{~}

\address{1. Center for Gravitational Physics and Geometry \\ Department
of Physics, The Pennsylvania State University \\ University Park, PA
16802, USA}

\address{2. Institute for Theoretical Physics,\\ University of
California, Santa Barbara, CA 93106, USA}

\maketitle

\begin{abstract}
A framework was recently introduced to generalize black hole mechanics
by replacing stationary event horizons with isolated horizons.  That
framework is significantly extended.  The extension is non-trivial in
that not only do the boundary conditions now allow the horizon to be
distorted and rotating, but also the subsequent analysis is based on
several new ingredients.  Specifically, although the overall strategy
is closely related to that in the previous work, the dynamical
variables, the action principle and the Hamiltonian framework are all
quite different.  More importantly, in the non-rotating case, the first
law is shown to arise as a \textit{necessary and sufficient condition
for the existence of a consistent Hamiltonian evolution}.  Somewhat
surprisingly, this consistency condition in turn leads to new
predictions even for \textit{static} black holes.  To complement the
previous work, the entire discussion is presented in terms of tetrads
and associated (real) Lorentz connections.
\end{abstract}

\section{Introduction}
\label{sec1}

The zeroth and first laws of black hole mechanics refer to equilibrium
situations and small departures therefrom.  The standard treatments
\cite{b,bch,carter,h,mh} restrict themselves to stationary space-times
admitting event horizons and small perturbations from stationarity.
While this simple idealization is a natural starting point, from
physical considerations it seems overly restrictive.  (See
\cite{ack,abf1} and especially \cite{abf2} for a detailed, critical
discussion.)  A framework, which is tailored to more realistic physical
situations was introduced in \cite{ack} and the zeroth and first laws
were extended to it in \cite{abf1,abf2,ac}.  This analysis generalizes
black hole mechanics in two directions.  First, the notion of event
horizons is replaced by that of `isolated horizons'. While the former
can only be defined retroactively, requiring access to the entire
space-time history, the latter can be defined quasi-locally. Second,
the underlying space-time need not admit \textit{any} Killing field;
isolated horizons need not be Killing horizons.  The static event
horizons normally used in black hole mechanics \cite{b,bch,carter,w2}
and the cosmological horizons in de Sitter space-times \cite{gh} are
all special cases of isolated horizons.  Furthermore, since space-times
can now admit gravitational and matter radiation, there is a large class of
other examples.

The framework developed in \cite{abf2} for generalizing black hole
mechanics was based on two restrictive assumptions.  First, only
undistorted, non-rotating horizons were considered. That is, the
boundary conditions used in \cite{abf2} implied that the
\textit{intrinsic} 2-metric of the horizon is spherically symmetric%
\footnote{However, it allowed space-times, such as the
Robinson-Trautman solutions, in which there is no space-time Killing
field whatsoever in any neighborhood of the isolated horizon \cite{c}.}
and that the imaginary part of the Weyl tensor component $\Psi_2$ ---
which encodes the angular momentum --- vanishes.  Second, while a
rather general class of matter fields was allowed, it was assumed that
the only relevant charges --- i.e., hair --- are the standard
electromagnetic ones.  The second assumption was weakened in
\cite{ac,cs} which allowed dilaton couplings and Yang-Mills fields. In
this paper, we allow for distortion \textit{and} more general matter
sources.  Distortion plays an important role in several astrophysical
situations, e.g., in problems involving black holes immersed in
external fields, or surrounded by matter rings, and especially in the
problem of black hole collisions.  Post-Newtonian considerations
suggest that, during black hole coalescence, individual horizons are
distorted due to the Coulomb attraction even in the regime in which
the black holes are sufficiently far from each other for the
gravitational radiation falling into their horizons to be negligible.
This phenomenon is also seen in numerical simulations.

The extensions \cite{ac,cs} which incorporated dilatonic and Yang-Mills
charges did not involve a significant generalization of the basic
framework developed in \cite{abf2}.  The present paper, on the other
hand, does.  We begin with substantially weaker boundary conditions
(formulated in terms of (real) tensor fields rather than the spinors
used in \cite{abf2}), and show that they imply constancy of surface
gravity (and electro-static potential) on the horizon.  This property
turns out to be \textit{necessary and sufficient} for the usual action
principle of tetrad gravity to continue to be valid in presence of
isolated horizons.  The action leads to a covariant phase space,
constructed from solutions to the field equations.  Ref \cite{abf2}, by
contrast, used the canonical phase space based on spinorial variables
which is tailored for quantization but which contains technical
complications that are unnecessary to the classical mechanics of
isolated horizons. \textit{Up to this point, distortion and rotation
are both incorporated}. However, in the last step, i.e., in the
discussion of the first law, we restrict ourselves to non-rotating
horizons. (Rotation is incorporated in \cite{abl2}.)

To formulate the first law one must first define the energy $E_\Delta$
associated with any isolated horizon $\Delta$. Since there can be
radiation in the spacetime outside isolated horizons, the ADM energy
$E_{\rm ADM}$ is not a good measure of $E_\Delta$ \cite{abf1,abf2}.
Instead, as in the work of Brown and York \cite{by}, the strategy is
to define the energy of the horizon using a Hamiltonian
framework. Experience with the phase space formulation of general
relativity suggests that, in the presence of boundaries, the
Hamiltonian $H_t$ generating time-translation along a suitable vector
field $t^a$ acquires surface terms.  The idea is to \textit{define
$E_\Delta$ as the surface term at $\Delta$ in the Hamiltonian}. The
key issue then is that of selecting the `appropriate' time-translation
$t^a$. Since one expects the volume term in the expression of $H_t$ to
be a linear combination of constraints and thus vanish when evaluated
on solutions, the problem reduces to that of specifying the boundary
values of $t^a$ (or, equivalently, of the lapse and shift fields).
The conditions at infinity are obvious and, in any case, will not
affect the surface term at $\Delta$.  Thus, we need to focus only on
the boundary value of $t^a$ on $\Delta$.

In the non-rotating case it is clear that, at the horizon, $t^a$ should
be proportional to the null normal to $\Delta$. However, our boundary
conditions do not select the null normal uniquely; there is a freedom
to rescale%
\footnote{This is not surprising since this freedom exists already on
Killing horizons. If the space-time is asymptotically flat and admits a
static Killing field globally, one can eliminate this freedom by
restricting oneself to that Killing field which is unit at infinity.
However, this strategy is not available if there is radiation in the
exterior region, or, as in the static solutions representing distorted
black holes, the metric fails to be asymptotically flat.}
the normal by a constant (on $\Delta$) which \textit{can vary from one
space-time to another}. This freedom is physically important because,
amongst other things, the surface gravity is sensitive to it.  Suppose
we fix this freedom by tying the boundary value of $t^a$ to fields on
the horizon, e.g., by demanding that surface gravity be a
\textit{specific} function of the area and charges.  We can then ask
whether the time-evolution generated by this $t^a$ preserves the
symplectic structure.  It turns out that the answer is not always in
the affirmative. On the horizon, the evolution vector field $t^a$ and
the electromagnetic potential $A_a$ have to be tied to the horizon
parameters appropriately.  These conditions impose a constraint on the
surface gravity $\kt$ and the electric potential $\Pt$.  Somewhat
surprisingly, the constraint is precisely the first law $\delta
E^t_{\Delta} = (\kt/8\pi G)\, \delta a_\Delta + \Pt\, \delta Q_\Delta$.
\textit{Thus the evolution defined by $t^a$ is
Hamiltonian if and only if the first law holds.}%
\footnote{In the undistorted context, while this role of the first law
was known to the authors of \cite{abf2}, its importance was not fully
appreciated.  The importance was noticed independently in \cite{bc} and
used effectively in \cite{cs} to extract physical information on
spherical black holes with Yang-Mills hair.}
In this sense, the first law is even more fundamental than it is
generally taken to be.  Conceptually, this is perhaps the most striking
feature of the present framework.

The requirement that the first law hold is not sufficient to fix $t^a$
uniquely.  Although every $t^a$ must be a null normal to $\Delta$, the
rescaling of $t^a$ from one spacetime to another can depend on the
horizon parameters and the first law does not fully determine this
parameter dependence.  There is an infinite family of
parameter-dependent vector fields $t^a$ each defining a consistent
Hamiltonian evolution and a horizon energy $E^t_{\Delta}$. By
contrast, at infinity all these vector fields must tend to a
time-translation of an universal flat metric, used in the construction
of the phase space.  Hence, there is only the familiar 3-parameter
freedom in the definition of $E^t_{\ADM}$, associated with the choice
of a rest-frame.  Furthermore, in any one space-time, we can eliminate
it by simply going to the rest frame and thus extract the total mass
$M_{\rm ADM}$ of the system.  Although not necessary for mechanics, it
is natural to ask if one can define a similar notion of mass of
isolated horizons.  The answer is in the affirmative in the
Einstein-Maxwell theory.  Let us require that $t^a$ should not only
lead to a consistent Hamiltonian evolution but agree, on static
solutions, with the static Killing field which is \textit{unit} at
infinity.  Then the horizon value of $t^a$ is uniquely determined for
\textit{all} space-times in the phase-space. There is a preferred
notion of time-translation, say $t^a_o$. We can set $M_\Delta =
E^{t_o}_{\Delta}$ and regard $M_\Delta$ as the mass of an isolated
horizon $\Delta$.  In the earlier work on non-distorted horizons
\cite{abf1,abf2,ac,cs}, the discussion of the first law was carried
out only in the context of these preferred evolution vector fields
$t^a_o$.  That derivation is more closely tied to the traditional
discussion of the laws in the static context but is not necessary from
the more general perspective of isolated horizons. Nonetheless, the
availability of a canonical definition of the mass $M_\Delta$ is
useful for other applications of this framework, e.g. to numerical
relativity.

The paper is organized as follows. In Section \ref{s2}, we specify the
boundary conditions defining general isolated horizons, allowing both
distortion and rotation. We explain the role of these conditions,
compare them with those used in \cite{ack,abf1,abf2,ac,cs} and work
out their consequences, including the zeroth law. In Section \ref{s3}
we introduce the Lagrangian framework based on tetrads and (real)
Lorentz connections and in Section \ref{s4}, the covariant phase
space. The first law is discussed in Section \ref{s5}. Upto this
point, the focus is on the Einstein-Maxwell theory (although
incorporation of the dilaton is straightforward). In Section \ref{s6}
we extend the framework to incorporate Yang-Mills fields. The horizon
mass is introduced in Section \ref{s7} and subtleties associated with
the dilaton and Yang-Mills fields are discussed. For the convenience
of readers who may not be familiar with distorted black holes,
Appendix A presents a variety of examples and, for convenience of
readers who work in the Newman-Penrose formalism, Appendix B
summarizes the structure of isolated horizons in that framework.

We have attempted to make this paper self-contained in terms of
methodology and technical details.  However, the motivation behind
isolated horizons and certain properties of our Hamiltonian are the
same as those discussed in detail in \cite{abf2}.  Since the inclusion
of distortion does not add anything substantial to these issues, we
have refrained from repeating that discussion in this paper.

\section{Structure of isolated horizons and the zeroth law}
\label{s2}

In this section, we will introduce the basic definitions of isolated
horizons and analyze their immediate consequences.  The definitions
will become progressively stronger.  However, even the strongest
boundary conditions are significantly weaker than requiring the horizon
to be a Killing horizon for a local Killing vector field. By proceeding
in steps, we will be able to keep track of the precise assumptions that
are needed to obtain various results. Also, the availability of a
hierarchy of definitions will be useful in other applications ---such
as numerical relativity and quantum gravity--- which lie beyond the
scope of the present paper.

Let us begin by introducing some notation.  Throughout this paper we
assume that all manifolds and fields under consideration are smooth.
Let $\man$ be a 4-manifold equipped with a metric $g_{ab}$ of signature
$(-,+,+,+)$. Let ${\Delta}$ be a null hypersurface of $(\man, g_{ab})$.
A \textit{future directed} null normal to $\Delta$ will be denoted by
$\l$.  (In this paper, the term `null normal' will always refer to a
future directed null normal.)  Let $q_{ab} \= g_{\pback{ab}}$ be the
degenerate intrinsic metric on $\Delta$.%
\footnote{Equalities which hold only at $\Delta$ will be denoted by
`${\hateq}$ ' and the pullback of a covariant index will be denoted by
an arrow under that index, e.g. $\omega_{\pback{a}}$ will denote the
pullback of the $1$-form $\omega_{a}$ to $\Delta$.}
A tensor $q^{ab}$ on $\Delta$ will be called an `inverse' of $q_{ab}$
if it satisfies $q^{ab}q_{ac}q_{bd} \= q_{cd}$.  Thus $q^{ab}$ is
unique only up to addition of terms of the form $\l^{(a} V^{b)}$ for
some vector field $V$ tangential to ${\Delta}$.  The expansion
$\theta_\ls$ of a specific null normal $\l$ is defined by $\theta_\ls =
q^{ab}\nabla_a \l_b$, where $\nabla_a$ is the derivative operator
compatible with $g_{ab}$. It is straightforward to check that
$\theta_\ls$ is independent of the choice of $q^{ab}$.  With this
structure at hand, we can now introduce our first definition.

\subsection{Non-expanding Horizons}
\label{s2.A}

\noindent\textbf{Definition 1}: A $3$-dimensional sub-manifold $\Delta$
of a space-time $(\man,g_{ab})$ is said to be a \textit{non-expanding
horizon} if it satisfies the following conditions:
\begin{itemize}
\item{(i)} $\Delta$ is topologically $S^2 \times \Rbar$ and null;

\item{(ii)} The expansion $\theta_{(\l)}$ of $\l$ vanishes on $\Delta$
for any null normal $\l$;

\item{(iii)}\label{eom} All equations of motion hold at $\Delta$ and
the stress-energy tensor $T_{ab}$ of matter fields at $\Delta$ is such
that $-T^{a}_{b}\l^{b}$ is future directed and causal for any future
directed null normal $\l$.
\end{itemize}
Note that if conditions (ii) and (iii) hold for one null normal $\l$
they hold for all.

The role of these conditions is as follows. The restriction on topology
is geared to the structure of horizons that result from gravitational
collapse. However, it can be weakened. One can retain the requirement
that the horizon have compact cross-sections but replace $S^2$ by a
manifold with higher genus. Our main analysis will extend to this case
in a straightforward manner. More generally, we can allow $\Delta$ to
have non-compact cross-sections, as for example in the case of certain
acceleration horizons. The results presented in this section, including
our derivation of the zeroth law will go through. However, since such
horizons extend to infinity, our Hamiltonian framework will have to be
modified appropriately.  Finally, one could envisage incorporation of
NUT charge. This extension would be even more subtle because, if all
fields are smooth, $\Delta$ would be topologically $S^3$ and $\l$ would
provide a Hopf fibration. In this case, $\Delta$ would not admit
\textit{any} cross-sections which are everywhere transverse to $\l$.
This extension will be discussed elsewhere.

Requirement (iii) is analogous to the dynamical conditions one imposes
at infinity.  While at infinity one requires that the metric (and other
fields) approach a specific solution to the field equations (the
`classical vacuum') , at the horizon we only ask that the field
equations be satisfied.  The energy condition involved is very weak; it
is implied by the (much stronger) dominant energy condition that is
typically used. Thus, the first and the last conditions are quite tame.

The key condition is (ii). It implies, in particular, that \textit{the
horizon area is constant `in time'} and thus incorporates the idea
that the horizon is isolated without having to assume the existence of
a Killing field. We will denote the area by $a_\Delta$ and refer to
$R_\Delta$ defined by $a_\Delta = 4 \pi R_\Delta^2$ as the
\textit{horizon radius}. All these conditions are satisfied on any
Killing horizon (with 2-sphere sections) if gravity is coupled to
physically reasonable matter (including perfect fluids, Klein-Gordon
fields, Maxwell fields possibly with dilatonic coupling, Yang-Mills
fields).

Although the conditions in the definition are quite weak, they have
surprisingly rich consequences. We will now discuss them in detail.
In some of this analysis it is convenient to use a null-tetrad and the
associated Newman-Penrose quantities (see Appendix B and references
therein). Given a specific null normal field $\l^a$ to $\Delta$, we
can introduce a complex null vector field $m^a$ tangential to $\Delta$
and a real, future directed null field $n^a$ transverse to $\Delta$ so
that the following relations hold: $n\cdot \l = -1, m\cdot\bar{m} =1$
and all other scalar products vanish. The quadruplet $(\l, n, m,
\bar{m})$ constitutes a null-tetrad. There is of course an infinite
number of null tetrads compatible with a given $\l$, related to one
another by restricted Lorentz rotations. Our conclusions will not be
sensitive to this gauge-freedom.

\medskip\noindent {(a)} \textit{Properties of} $\l$: Since $\l^{a}$
is a null normal to $\Delta$, it is automatically twist free and
geodesic. We will denote the acceleration of $\l^a$ by $\kappa_{\ls}$
\begin{equation} \label{kappa}
\l^{a}\grad_{a}\l^{b} \= \kappa_{\ls}\l^{b} \, .
\end{equation}
Note that the acceleration is a property not of the horizon $\Delta$
itself, but of a specific null normal to it: if we replace $\l$ by
$\l^\prime = f \l$, then the acceleration changes via
$\kappa_{(\l^\prime)} = f\kappa_{\ls} + {\cal L}_\l f$.

Since the twist of $\l$ vanishes, the Raychaudhuri equation implies:
\[
\Lie_{\l}\theta_{\ls} \hateq \kappa_{\ls}\theta_{\ls} - \half
\theta_{\ls}^{2} - {\sigma}\bar{\sigma} - R_{ab}\l^{a}\l^{b}
\]
where $\sigma = m^a m^b \nabla_a \l_b$ is the shear of $\l$ in the
given null tetrad. Since $\theta_{\ls}$ vanishes on $\Delta$, we
conclude: $\sigma \, \overline{\sigma} + R_{ab}\l^a\l^b \= 0$. The
condition on the stress-energy tensor ensures that $R_{ab}\l^a \l^b =
8\pi G T_{ab}\l^a\l^b$ is non-negative on $\Delta$.  Hence, we
conclude:
\begin{equation}\label{rayres}
\sigma \= 0, \quad {\rm and} \quad R_{ab}{\l}^a{\l}^b \=0.
\end{equation}
Thus, in particular, every null normal $\l$ is free of expansion, twist
and shear.

\medskip\noindent {(b)} \textit{Conditions on the Ricci tensor}: The
second equation in (\ref{rayres}) implies that the vector
$-{R^{a}}_{b}\l^{b}$ is tangential to $\Delta$.  The energy condition
and the field equations imply this vector must also be future causal.
This means that $R^a{}_{b}\l^{b}$ must be proportional to $\l^{a}$ and
hence, ${R_{\pback{a}b}}\l^{b} = 0$. In the Newman-Penrose formalism
this condition translates to:
\begin{equation}\label{phi00and10}
    \Phi_{00} = \half R_{ab}\l^{a}\l^{b} \= 0
    \quad {\rm and} \quad
    \Phi_{01} = \bar{\Phi}_{10} = \half R_{ab}\l^{a}m^{b} \= 0 .
\end{equation}
Since this statement is equivalent to $R_{\pback{a}b}\l^b = 0$, it is
gauge invariant, i.e. it does not depend upon the specific choice of
null normal $\l$ and $m$.

\medskip\noindent  {(c)} \textit{A natural connection 1-form on}
$\Delta$: Since $\l$ is expansion, shear and twist free, there exists a
one-form $\omega_{a}$ intrinsic to $\Delta$ such that
\begin{equation}\label{omegadefn}
\grad_{\!\pback{a}}\l^{b} \= \omega_{a}\l^{b}
\end{equation}
which in turn implies
\[
{\cal L}_\l q_{ab} \= 2\pback{\grad_{a}\l_{b}} \= 0\, .
\]
Thus, every null normal $\l$ is a `Killing field' of the degenerate
metric on $\Delta$. Furthermore, we will now show that
\begin{equation}
\label{2epsilon} {}^2\!\epsilon := i m\wedge \bar{m}
\end{equation}
is also invariantly defined.  Since ${\cal L}_\l q_{ab}= 0$, the space
$\S$ of integral curves of $\l$ is naturally equipped with a
non-degenerate metric $\underline{q}_{ab}$ (so that $q_{ab}$ on
$\Delta$ is the pull-back of $\underline{q}_{ab}$). Denote by
$\underline{\epsilon}_{ab}$ the unique (up to orientation) unit
alternating tensor on $({\S},\underline{q}_{ab})$. ${}^2\!\epsilon$ is
the pull-back to $\Delta$ of $\underline\epsilon$. Although $\ell$ is a
`Killing field' of the intrinsic horizon geometry, the space-time
metric $g_{ab}$ need \textit{not} admit a Killing field in any
neighborhood of $\Delta$.  Robinson-Trautman metrics \cite{c} and the
Kastor-Traschen solutions \cite{kt} provide explicit examples of this
type.

The 1-form $\omega$ will play an important role throughout this paper.
It has an interesting geometrical interpretation. We can regard
$\omega$ as a connection on the line bundle $T\Delta^\perp$ over
$\Delta$ whose fibers are the 1-dimensional null normals to $\Delta$.
Under the rescalings $\l \mapsto \tilde{\l} = f\l$, of the null normal
$\l$, it transforms via:
\be \label{omegatrans} \omega_{a}\mapsto \tilde{\omega}_{a} =
\omega_{a} + \grad_{\pback{a}}\ln f. \ee

\medskip\noindent {(d)} \textit{Induced Connection on} $\Delta$: Each
metric submanifold $M$ of $\man$ admits a natural connection --- one
which is torsion-free and compatible with the induced metric on $M$.
This connection is also canonically induced by the space-time
connection $\grad$. However, since the induced metric $q_{ab}$ on
$\Delta$ is degenerate, there exist infinitely many connections
compatible with it. A general null sub-manifold inherits a unique
(torsion-free) derivative operator $\D$ from $\grad$ if and only if its
null normal $\l$ satisfies $\pback{\grad_{a}\l_{b}}=0$.  Therefore, the
conditions imposed in Definition 1 guarantee that every non-expanding
horizon has a unique intrinsic derivative operator $\D$.  The action of
$\D$ on a vector field $X^{a}$ tangent to $\Delta$ and on a 1-form
$\eta_{a}$ intrinsic to $\Delta$ is given by
\[
\D_{a}X^{b} \= \pback{\grad_{a}}\tilde{X}^b \qquad \mbox{and} \qquad
\D_{a}\eta_{b} \= \pback{\grad_{a}\tilde{\eta}_{b}} \,\, .
\]
where $\tilde{X}^b$ and $\tilde{\eta}_{b}$ are arbitrary extensions of
$X^b$ and $\eta_b$ to the full space-time manifold $\man$. It is easy
to show that $\D$ is independent of the extensions.

The 1-form $\omega$ captures only part of the information in $\D$. The
full connection $\D$ on $\Delta$ plays an important role in extracting
physics in the strong field regime near $\Delta$ \cite{abl1}. However,
it is not essential to the discussion of isolated horizon mechanics.

\medskip\noindent  {(e)} \textit{Conditions on the Weyl tensor}: Let
us begin with the definition of the Riemann tensor,
$[\grad_{a}\grad_{b} - \grad_{b}\grad_{a}]X^{c} = -2{R_{abd}}^{c}
X^{d}$.  If we set $X^c = \ell^c$ and pull back the indices $a$ and
$b$, then using (\ref{omegadefn}), we obtain:
\begin{equation}\label{dw}
 [\D_{a}\omega_{b}-\D_{b}\omega_{a}]\l^{c} \= -2R_{\pback{abd}}{}^{c}\l^{d}
 \hateq - 2C_{\pback{abd}}{}^{c}\l^{d}
\end{equation}
where ${C_{abc}}^d$ is the Weyl tensor. The last equality follows from
$R_{\pback{ab}}\l^{b} \hateq 0$. Thus, if $v$ is any $1$-form on $\Delta$
satisfying $v \cdot \l \hateq 0$, contracting the previous equation
with $v_{c}$ we get
\[ C_{\pback{abd}}{}^cv_{c}\l^{d} \hateq 0 \, .\]
Let us choose a null tetrad and set $v$ to be $m$ or $\bar{m}$. Then
\begin{equation}\label{psi0psi1}
\Psi_{0} := C_{abcd}\l^{a}m^{b}\l^{c}m^{d} \= 0 \qquad \mbox{and}
\qquad \Psi_{1} := C_{abcd}\l^{a}m^{b}\l^{c}n^{d} =
C_{abcd}\l^{a}m^{b}\bar{m}^{c}m^{d}\= 0\, ,
\end{equation}
where we have used the trace-free property of the Weyl tensor in the
second equation. It is also clear that equations (\ref{psi0psi1}) are
independent of which null normal $\l$, and vector fields $m$ and
$\bar{m}$ we choose to construct the null tetrad; equation
(\ref{psi0psi1}) is gauge invariant.

\medskip\noindent  {(f)} \textit{Curvature of} $\omega$: Let us
contract (\ref{dw}) with $n_{c}$ and use $\l^{a}n_{a} = -1$.  Then we
have:
\begin{equation}\label{derivomega}
2\D_{[a}\omega_{b]} \hateq {C_{\pback{abd}}}^{c}\l^{d}n_{c} \hateq
C_{\pback{abc}d}\l^{c}n^{d} \, .
\end{equation}
Expanding the Weyl tensor in terms of the $\Psi$'s, one obtains
\begin{eqnarray}
 C_{abcd}\l^{c}n^{d} &\=& 4(\RePt{\Psi_{2}})n_{[a}l_{b]} +
 2\Psi_{3}\l_{[a}m_{b]} + 2\bar{\Psi}_{3}\l_{[a}\bar{m}_{b]}
 \nonumber \\
&& -2\bar{\Psi}_{1}n_{[a}m_{b]} - 2\Psi_{1} n_{[a}\bar{m}_{b]} +
4i(\ImPt{\Psi_{2}})m_{[a}\bar{m}_{b]} \, .
\end{eqnarray}
where
\begin{equation}\label{psi2psi3}
\Psi_{2} :\= C_{abcd}\l^{a}m^{b}\bar{m}^c n^{d}  \qquad
\mbox{and}\qquad \Psi_{3} :\= C_{abcd}\l^{a}n^{b}\bar{m}^{c}n^{d} \, .
\end{equation}
Substituting this expression into (\ref{derivomega}), pulling back on
the two free indices and taking into account (\ref{psi0psi1}) and
(\ref{2epsilon}), we obtain
\begin{equation} \label{omegacurv}
d\omega \hateq 2(\ImPt{\Psi_{2}})\, \twoeps \, .
\end{equation}
This relation will play an important role in what follows. Note that,
because $\Psi_0$ and $\Psi_1$ vanish on $\Delta$, $\Psi_2$ is gauge
invariant.
\medskip

\medskip\noindent
{\sl Remark}: It is interesting to compare the structure of $\Delta$
with that of null-infinity $\I$ (in the usual conformal gauge, in which
the conformal factor is chosen such that the null-normal to $\I$ is
divergence-free).  Both are null surfaces and can be regarded as `line
bundles' over a base space $\S$ of the integral curves of null normals.
(For brevity, we will ignore a caveat concerning completeness of
fibers.) The null normals are Killing fields of the intrinsic
degenerate metric so that this metric is the pull-back to the 3-surface
of a positive-definite metric on $\S$.  In both cases, the space-time
connection induces an intrinsic connection on the 3-surface \cite{as}.
These connections capture physically important information.  However,
there are a number of differences as well. Since $\I$ is constructed by
a conformal completion, the conformal freedom permeates all geometric
structures at $\I$.  In particular, given a physical space-time, the
intrinsic metric and the derivative operator are known only up to
conformal transformations.  On the other hand, since $\I$ is at
infinity, in some ways its structure is both more rigid and simpler.
First, without loss of generality, we can assume that the metric on
$\S$ is a 2-sphere metric; the issue of distortion is physically
irrelevant at $\I$. Second, the Weyl tensor vanishes identically at
$\I$ and the curvature of the intrinsic connection captures non-trivial
information about the next order space-time curvature.  By contrast, at
$\Delta$ only four components of the Weyl curvature vanish and four
other components are coded in the curvature of the intrinsic connection
on $\Delta$. In spite of these differences, one can carry over some
techniques from null infinity to extract physical information about
isolated horizons. In particular, using the analogs of techniques which
have been successful at $\I$, one can introduce preferred cross
sections of and Bondi-type expansions near $\Delta$ \cite{abl1}.

This concludes our analysis of the consequences of the boundary
conditions defining non-expanding horizons. Note that, even though $\l$
is a Killing field for the intrinsic, degenerate metric $q_{ab}$ on
$\Delta$, it is not an infinitesimal symmetry for other geometrical
fields such as the intrinsic connection $\D$ or components of the
curvature tensor. In the next sub-section, we will make the structure
more rigid by suitably restricting the choice of $\l$.

\subsection{ Weakly Isolated horizons}
\label{s2.B}

The time-independence of the intrinsic metric $q_{ab}$ captures the
idea that $\Delta$ is isolated in a suitable sense.  While this
condition has rich consequences, the resulting structure is still not
sufficient for physical applications.  In particular, since $\l$ can be
rescaled by an arbitrary positive definite function, the acceleration
$\kappa_{\ls}$ is not necessarily constant on $\Delta$. Therefore, we
need to impose additional restrictions on the physical fields at
$\Delta$ to establish the zeroth law.  Since $\ell$ is already a
symmetry of the intrinsic metric, it is natural to require it also be a
symmetry of the `extrinsic curvature'. However, the standard definition
of the extrinsic curvature is not applicable to null surfaces.
Nonetheless, given a null normal $\ell$, we can construct a tensor
field $K_a{}^b := \D_a \l^b$, defined intrinsically on $\Delta$, which
can be thought of as the analogue of the extrinsic
curvature.%
\footnote{We are grateful to Thibault Damour for pointing out that
$K_a{}^b$ is called the Weingarten map and is analogous to extrinsic
curvature. This comment suggested the above motivation for our
condition on the connection 1-form $\omega$. For an alternate, and in a
sense weaker, condition see the remark at the end of Section
\ref{s2.C}.  From the viewpoint of intrinsic structures on $\Delta$
discussed in Section \ref{s2.A}, it is perhaps more natural to ask that
$\l$ be a symmetry of the \textit{full} intrinsic connection $\D$ (see
section \ref{s2.C} and \cite{abl1}.)  However, this stronger condition
is not necessary for the laws of mechanics discussed here.}
Indeed, on a metric sub-manifold, if we replace $\l$ by the
\textit{unit} normal, $K_a{}^b$ is precisely the extrinsic curvature.
It is then natural to demand that, on an isolated horizon, $K_a{}^b$
also be time-independent: ${\cal L}_{\l} K_a{}^b \= 0$. As a
consequence of (\ref{omegadefn}), this is equivalent to imposing ${\cal
L}_\l \omega \= 0$.

Let us examine the above condition.  As we will show at the end of this
section, given a non-expanding horizon we can always find a null normal
$\ell^a$ which satisfies ${\cal L}_\l \omega \= 0$. The behavior of
this condition under rescalings of $\ell$ is complicated by the fact
that $\omega$ itself depends upon the choice of null normal (see
equation (\ref{omegatrans})).  However, under a constant rescaling $\l
\mapsto \tilde\l = c\l$, the connection 1-form $\omega$ is unchanged.
Therefore, if $\ell$ satisfies the condition ${\cal L}_{\l} \omega \=
0$, so does any $\tilde{\ell}$ related to $\ell$ by constant rescaling.
This suggests we introduce an equivalence relation: \textsl{Two
future-directed null normals $\l$ and $\tilde\l$ belong to the same
equivalence class $[\l]$ if and only if $\tilde{\l} = c\l$ for some
positive constant $c$.}

The above considerations lead us to the following definition:

\medskip
\noindent \textbf{Definition 2}: A weakly isolated horizon $(\Delta,
[\l])$ consists of a non-expanding horizon $\Delta$, equipped with an
equivalence class $[\l]$ of null normals to it satisfying
\be \label{i} {\cal L}_{\l} \omega \=0\,\,\, {\hbox{\rm for all}}\,\,
 \l \in [\l].\ee
\noindent As pointed out above, if this last equation holds for one
$\l$, it holds for all $\l$ in $[\l]$.

A Killing horizon (with 2-sphere cross-sections) is automatically a
weakly isolated horizon, (provided the matter fields satisfy the energy
condition of Definition 1).  Given a non-expanding horizon $\Delta$,
one can always find an equivalence class $[\l ]$ of null-normals such
that $(\Delta, [\l ])$ is a weakly isolated horizon. However, condition
(\ref{i}) does not by itself single out the appropriate equivalence
class $[\l]$. As indicated in Section \ref{s2.C}, one \textit{can}
further strengthen the boundary conditions and provide a specific
prescription to select the equivalence class $[\l ]$ uniquely. However,
for mechanics of isolated horizons, these extra steps are unnecessary.
In particular, our analysis will not depend on how the equivalence
class $[\l ]$ is chosen. The adverb `weakly' in Definition 2 emphasizes
this point.

The condition (\ref{i}) has several consequences which are relevant for
this paper.

\medskip\noindent  {(a)} \textit{Surface gravity}: In the case of
Killing horizons $\Delta_{\rm K}$, surface gravity is defined as the
acceleration of the Killing field $\xi$ normal to $\Delta_{\rm K}$.
However, if $\Delta_{\rm K}$ is a Killing horizon for $\xi$, it is
also a Killing horizon for $c\,\xi$ for any positive constant $c$.
Hence, surface gravity is not an intrinsic property of $\Delta_{\rm
K}$, but depends also on the choice of a specific Killing field $\xi$.
(Of course the result that the surface gravity is constant on
$\Delta_{\rm K}$ is insensitive to this rescaling freedom.) In
asymptotically flat space-times admitting global Killing fields, this
ambiguity is generally resolved by selecting a preferred normalization
in terms of the structure at infinity. For example, in the static case,
one requires the Killing field $\xi$ to be unit at infinity. However,
in absence of a \textit{global} Killing field or asymptotic flatness,
this strategy does not work and we simply have to accept the constant
rescaling freedom in the definition of surface gravity. In the context
of isolated horizons, then, it is natural to keep this freedom.

A weakly isolated horizon is similarly equipped with a preferred family
$[\l]$ of null normals, unique up to constant rescalings. Therefore, it
is natural to interpret $\kappa_{\ls}$ as the surface gravity
associated with $\l$. Under the permissible rescalings $\l \mapsto
\tilde\l = c \l$, the surface gravity transforms via:
$\kappa_{(\tilde{\l})} = c \kappa_{\ls}$. Thus, while $\omega$ is
insensitive to the rescaling freedom in $[\l]$, $\kappa_\ls$ captures
this freedom fully.  One can, if necessary, select a specific $\l$ in
$[\l]$ by demanding that $\kappa_\ls$ be a specific function of the
horizon parameters which are insensitive to this freedom, e.g., by
setting $\kappa_\ls = 1/2R_\Delta$, where $R_\Delta$ is the horizon
radius (related to the horizon area $a_\Delta$ via $a_\Delta = 4\pi
R^2_\Delta$).

\medskip\noindent {(b)} \textit{Zeroth law}:  The boundary conditions
of Definition 2 allow us to define surface gravity $\kappa_{(\ell)}$ of
a weakly isolated horizon $(\Delta, [\l])$.  We will now show that the
surface gravity is constant on $\Delta$.  In other words, the zeroth
law holds for weakly isolated horizons.

Recall from (\ref{omegacurv}) that on a non-expanding horizon,
$d\omega \= 2\, \ImPt{\Psi_2} \,{}^2\!\epsilon$ for any choice of null
normal $\l$.  Since $\twoeps$ is the pull-back to $\Delta$ of the
alternating tensor $\underline\epsilon$ on the space $\S$ (of orbits
of $\l$), clearly $\ell \into \twoeps \= 0$.  Therefore,
\[
\ell \into d\omega \hateq 0
\]
for every null normal $\ell$. In particular, on a weakly isolated
horizon this equation holds for any $\l \in [\l]$.  Moreover, each of
these restricted null normals also satisfies
\[
    0 \hateq {\cal L}_\l \omega
        \hateq d(\ell \cdot \omega) + \ell \into d\omega
\]
Hence, we conclude:
\[ d(\ell \cdot \omega)  \hateq d(\kappa_{\ls}) \hateq 0\, , \]
where we have used the definition (\ref{kappa}) of $\kappa_{\ls}$.
Thus, surface gravity is constant on $\Delta$.

Although this proof of the zeroth law appears extremely simple, the
argument is not as trivial as it might first appear since we have used
a number of consequences of the boundary conditions derived in section
\ref{s2.A}. In contrast to earlier derivations \cite{bch,w2} we do not
require the presence of a Killing field even in a neighborhood of
$\Delta$. Therefore the proof applies also to space-times such as the
Robinson-Trautman solutions \cite{c} which do not admit a Killing
field. Also, $\Delta$ need not be `complete' --- it may be of finite
affine length with respect to any $\l$ --- and may not admit the analog
of a `bifurcate surface' on which the Killing field vanishes. Finally,
the field equations are used rather weakly; we only need to assume that
$-(R^a{}_b - \textstyle{1\over 2} R \delta^a{}_b)\l^b$ is a future
directed causal vector.

Surface gravity does not have a definite value on a weakly isolated
horizon.  The value of $\kappa_{(\ell)}$ depends upon the choice of
null normal $\ell \in [\ell]$. Since all the normals $\l$ to $\Delta$
are future directed, the rescaling constant $c$ is necessarily
positive. Therefore, if the surface gravity is non-zero (respectively,
zero) with respect to one $\l$, it is non-zero (respectively, zero)
with respect to any other $\l \in [\l]$.  This rescaling freedom is the
same as the one discussed above in the context of Killing horizons.

We will conclude this sub-section with three remarks.

i) \textsl{Freedom in the choice of} $[\l ]$: Given a non-expanding
horizon $\Delta$, it is natural to ask if one can always select an
equivalence class $[\l ]$ of null normals such that $(\Delta, [\l ])$
is a weakly isolated horizon. As indicated earlier in this section, the
answer is in the affirmative and, furthermore, there is a considerable
freedom in the choice of $[\l ]$. Let us examine this issue in some
detail.

Since $\l \into d \omega \= 0$ for any null normal $\l$ to a
non-expanding horizon, it follows that a null normal $\l$ satisfies
${\cal L}_\l \omega \= 0$ if and only if $d\kappa_\ls \= 0$. Thus, to
find a family $[\l]$ required in the definition of weakly isolated
horizons, it is necessary and sufficient to find a null normal $\l$ for
which the surface gravity is constant.  On a non-expanding horizon, the
surface gravity transforms as follows.
\[ {\rm If}\,\,
\l \mapsto \tilde{\l} \= f \l  ,\quad {\rm  then} \quad
\kappa_{(\tilde{\l})} \= f \kappa_\ls + {\cal L}_\l f .\]
Hence, starting with any $\l$, we can simply solve for $f$ by
requiring that $\kappa_{(\tilde{l})}$ be constant on $\Delta$. The
solution is not unique.  If $\kappa_\l$ is constant, given any
non-zero function $g$ satisfying ${\cal L}_\l g \= 0$ and a constant
$\tilde\kappa$, let us set
\[ f \= g\, e^{-\kappa_\ls\, v} + {\tilde\kappa \over \kappa_\ls} \]
where $v$ satisfies ${\cal L}_{\l} v \= 1$.  Then,  we obtain an
$\tilde\l \not\in [\l]$ for which $\kappa_{(\tilde{\l})} \=
\tilde\kappa$.  This is the only freedom if both $\kappa_\l$ and
$\kappa_{(\tilde\l)}$ are to be constant.  Thus, each non-expanding
horizon gives rise to an infinite family of weakly isolated horizons.
Put differently, although one can easily obtain weakly isolated
horizons from non-expanding ones by choosing appropriate null normals
$[\l]$, a specific weakly isolated horizon carries much more
information than the non-expanding horizon it comes from.  At the end
of Section \ref{s2.C}, we will indicate how one can further strengthen
the boundary conditions to give a prescription for selecting a specific
$[\l ]$.  However, the analysis of this paper does \textit{not} depend
on how this selection is made.

ii) How does Definition 2 compare with that used in the undistorted,
non-rotating case?  As one would expect, the definition given in
\cite{ack,abf1,abf2} is significantly stronger. Furthermore, it was
tied to a foliation from the beginning.  More precisely, it assumed
that there exists a foliation to which $\omega$ is normal, with $\omega
= -\kappa n$, and it required that the expansion $\RePt\mu$ of the null
normal $n$ to the leaves of the foliation be constant on $\Delta$ (see
appendix \ref{appb} for definitions of the NP spin coefficients).
Although it was shown that the foliation is unique if it exists, the
heavy reliance on the foliation right from the beginning made that
definition less elegant and its invariant content less transparent.
Also, since we now allow $d\omega$ to be non-zero and impose no
conditions on $\RePt\mu$, we can now incorporate rotation and
distortion.

iii) \textsl{Alternate boundary conditions}: In the definition of
isolated horizons, we required ${\cal L}_\l \omega \= 0$, which in
particular implies ${\cal L}_\l \kappa_\ls \equiv {\cal L}_\l (\ell
\cdot \omega) \= 0$.  Thus, the definition itself assured us that
$\kappa_\ls$ is `time independent' and to prove the zeroth law we had
to show that it is also independent of `angles'.  Could we have used
another definition in which the `time dependence' of $\kappa_\ls$ was
not explicitly required but followed from other conditions?  The answer
is in the affirmative: In place of ${\cal L}_\l \omega \=0$, we could
have required that $\Delta$ admit a foliation on which the expansion
$\RePt\mu$ of $n$ and the Newman-Penrose spin coefficient $\pi$ which
carries the angular momentum information are `time independent'
\cite{abl2}.  Then the field equations would have \textit{implied} that
$\kappa_\ls$ is time-independent.  Furthermore, all results of this
paper go through (and were in fact first obtained) with these modified
boundary conditions.  Note however that the new condition is neither
weaker nor stronger than the one we used.  Both require that $\pi$ be
time independent.  In addition, the present definition of isolated
horizons requires that $\kappa_\ls$ be time independent while the
alternative definition would have required, instead, that $\RePt\mu$ be
time independent.  In this paper, we chose the present definition
because it can be stated without reference to a foliation.

\subsection{Electromagnetic Field}
\label{s2.D}

We shall now describe the form of the electromagnetic field at an
isolated horizon and introduce a partial gauge fixing at the horizon
which will allow us to introduce the notion of an electric potential.
In the next three sections --- where we discuss the action, phase space
and first law --- we will assume that the only matter field present
\textit{at} the isolated horizon are Maxwell fields. However, as our
discussion will make it clear, this restriction is made primarily for
simplicity. The overall framework is rather general and can accommodate
matter for which there exists a well-defined action principle and a
(covariant) Hamiltonian framework. In particular, in Section \ref{s6},
we describe how to extend the formalism to include Yang Mills Fields.

The isolated horizon boundary conditions restrict matter primarily
through conditions on the stress-energy tensor $T_{ab}$. Let us begin
with $T_{ab}\ell^a \ell^b \hateq 0$, a direct consequence of the
boundary conditions and Raychaudhuri equation. (This restriction arises
due to the fact that $\Delta$ is a non-expanding horizon; the
subsequent stronger boundary conditions do not further constrain
$\emF$). Although this condition is weak, it turns out to have
interesting consequences on the form of the electromagnetic field,
$\emF$, at $\Delta$.  The stress-energy tensor for electromagnetism is
given in terms of the field strength $\emF$ as
\begin{equation}\label{emt}
\emT_{ab} = \frac{1}{4\pi}[\emF_{ac} {\emF_b}^c
    - \frac{1}{4}g_{ab} \emF_{cd}\emF^{cd}].
\end{equation}
Let us contract this expression with $\ell^a \ell^b$ and examine the
consequences for $\emF$.  This gives
\begin{equation}\label{emtll}
0 \hateq \emT_{ab}\ell^a \ell^b \hateq \mid\! \ell^a m^b \emF_{ab}
\!\mid ^2 \, ,
\end{equation}
where, to obtain the last expression, we have used the anti-symmetry of
$\emF$ and the fact that the metric at the horizon can be expressed in
terms of a null-tetrad as $g_{ab}= -2\ell_{(a}n_{b)}
+2m_{(a}\bar{m}_{b)}$.  An immediate consequence of (\ref{emtll}) is
that
\begin{equation}\label{pbackf}
\pback{\ell^a \emF_{ab}} \hateq 0.
\end{equation}
In order to  obtain a similar expression for $\dual \emF$ recall that
the stress energy tensor can be rewritten as $\emT_{ab} =
-\frac{1}{4\pi} [\dual\emF_{ac} {\dual\emF_b}^c - \frac{1}{4}g_{ab}
\dual\emF_{cd}\dual\emF^{cd}]$.  Applying the same argument which led
to (\ref{pbackf}), we obtain a similar restriction on $\dual \emF$,
namely
\begin{equation}\label{pbackstarf}
\pback{\ell^a  \dual{\emF}_{ab}} \hateq 0.
\end{equation}
These two restrictions tell us there is no flux of electromagnetic
radiation across the horizon.

It is straightforward to show that (\ref{pbackf}), (\ref{pbackstarf})
and the form of the metric at $\Delta$ place further restrictions on
${\emT}_{ab}$:
\[
\begin{array}{rcl@{\hspace{5mm}}rcl}
    \emT_{ab} \ell^a m^b &\hateq& 0 &
    \emT_{ab} \ell^a \bar{m}^b &\hateq& 0 \\[1.5mm]
    \emT_{ab} m^a m^b &\hateq& 0 &
    \emT_{ab} \bar{m}^a \bar{m}^b &\hateq& 0.
\end{array}
\]

The first two equations contain no new information since we already
knew from general arguments (see equation (\ref{phi00and10})), that
$\Phi_{10}\hateq 0$ and $\Phi_{01} \hateq 0$. However, the last two
equations do place further restrictions on the stress energy tensor.
Since the equations of motion are enforced at the boundary, we see
immediately that they are equivalent to further restricting the Ricci
tensor at the horizon by requiring:
\begin{equation}\label{phi20}
\Phi_{02} := \half R_{ab} {m}^a {m}^b \hateq 0 \qquad \mbox{and} \qquad
\Phi_{20} := \half R_{ab} \bar{m}^a \bar{m}^b \hateq 0 \, .
\end{equation}
This result need not hold for general matter fields; it relies on the
properties of the electromagnetic stress-energy tensor (\ref{emt}).

The next task is to define the electric and magnetic charges of the
horizon.  Since the horizon is an inner boundary of spacetime, the
normal to a 2-sphere cross section of the horizon will naturally be
inward pointing. Bearing this in mind, we define the electric and
magnetic charges of the horizon as
\begin{equation}\label{charge}
Q_{\Delta} :\hateq -\frac{1}{4\pi} \oint_{S_{\Delta}} \dual \emF \qquad
\mbox{and} \qquad P_{\Delta} :\hateq -\frac{1}{4\pi} \oint_{S_{\Delta}}
\emF \, .
\end{equation}
For these definitions to be meaningful, the values of $Q_{\Delta}$ and
$P_{\Delta}$ should be independent of the cross section of the horizon
$S_{\Delta}$.  We will now show that $\Delta$ being a non-expanding
horizon guarantees this is the case.  Let us first evaluate
\[
\Lie_{\ell}\pback{\emF} \hateq
    \ell \into \pback{d\emF} + \pback{d(\ell \into \emF)}.
\]
The first term on the right hand side vanishes due to Maxwell's
equations on $\Delta$, while the second term is zero due to the
previous restriction on $\emF$, (\ref{pbackf}). Therefore we conclude
that $\pback{\emF}$ is Lie dragged by $\ell$.  An identical argument
for $\dual \emF$ leads to the analogous conclusion.  Therefore we
obtain
\begin{equation}\label{lief}
\Lie_{\ell}\pback{\emF} \hateq 0 \qquad \mbox{and} \qquad
\Lie_{\ell}\pback{\dual\emF} \hateq 0 \, .
\end{equation}
This result, along with (\ref{pbackf}) and (\ref{pbackstarf}),
guarantees that $Q_{\Delta}$ and $P_{\Delta}$ are independent of the
choice of cross section $S_{\Delta}$ of the horizon.  Note that this
result was obtained using only the boundary conditions; equations of
motion in the bulk are not needed.

Finally, let us examine the remaining freedom in the electromagnetic
field.  The boundary conditions do not restrict $\emF_{ab} n^a m^b$ and
$\dual\emF_{ab} n^a m^b$ at all.  These components describe the
electromagnetic radiation flowing along the horizon.  Therefore,
isolated horizon boundary conditions allow electromagnetic radiation
arbitrarily close to ---and even at--- the horizon, provided none
crosses it.

So far we have confined ourselves to the field strengths $\emF$ and
$\dual\emF$.  However, in the action principle and the Hamiltonian
framework we have to consider also the Maxwell potential $\emA$.  Now,
if the magnetic charge is non-zero, either one has to allow `wire
singularities' in the vector potentials or regard $\emA$ as a
connection on a non-trivial $U(1)$-bundle.  (If we regard it as a
connection on a $\Rbar^+$-bundle, the magnetic charge is necessarily
zero.) Since we wish to deal only with smooth fields, we will not allow
`wire singularities' in the potentials.  If we work with bundles, the
magnetic charge is quantized whence the space of histories has several
disconnected components. Thus, in the first law, we will not be able to
consider variations $\delta$ of fields with $\delta P \not=0$. As far
as mechanics of isolated horizons is concerned, there is essentially no
loss of generality if we restrict ourselves to the case $P_\Delta = 0$.
Therefore, in the next three sections, while working with Maxwell
fields, we will do so.  As usual, our \textit{final} results can be
formally extended to the case of non-vanishing magnetic charge by
performing a duality rotation on $\emF$.  (As discussed in Section
\ref{s6}, the situation is rather different in the case of Yang-Mills
fields.)

Recall that the first law in the Einstein-Maxwell case involves the
electro-static potential $\Phi$.  In static space-times, one typically
sets $\Phi = - \xi^a\emA_a$ where $\xi$ is the static Killing field and
the gauge is chosen such that the vector potential $\emA$ tends to zero
at infinity and satisfies ${\cal L}_\xi \emA =0$ \textit{everywhere in
space-time}.  We now need a strategy to define the electric potential
$\Phi$ \textit{without} reference to a Killing field or infinity. To
this end, we introduce the following definition:

\medskip\noindent\textbf{Definition 3}: The electromagnetic potential
$\emA$ will be said to be in a \textit{gauge adapted to the weakly
isolated horizon} $(\Delta, [\l])$ if it satisfies
\begin{equation}\label{lieema}
\Lie_{\ell} \pback{\emA} \hateq 0.
\end{equation}

\medskip\noindent Mathematically, this restriction is analogous to
this condition $\Lie_{\ell} \omega \hateq 0$ imposed on the
gravitational field in Definition 2.  However, while the condition on
$\omega$ is a physical restriction on the form of the gravitational
field at $\Delta$, the condition on $\emA$ is a gauge choice; it can
always be imposed without physically constraining the electromagnetic
field strength.

Given an electromagnetic potential $\emA$ in a gauge adapted to
$(\Delta, [\l])$, we can now define the scalar potential $\Phi_{(\l)}$
\textit{at} the horizon in an obvious fashion:
\[
\Phi_{(\ell)}:\hateq - \ell \cdot \emA.
\]
The key question now is whether our boundary conditions are strong
enough to ensure that $\Phi_{(\l)}$ is constant on $\Delta$. Only then
can we hope to use this notion of the scalar potential in the first
law.  Note that this question is rather similar to the one we asked of
surface gravity $\kappa_\ls$ in Section \ref{s2.B}. By using arguments
completely analogous to those that led us to the zeroth law, we will
now show that the answer to the present question is also in the
affirmative.  In a gauge adapted to the horizon,
\[
    0 \hateq \Lie_{\ell} \pback{\emA}
    \hateq \pback{\ell \into \emF} - \pback{d\Phi}_{(\ell)}.
\]
As we saw above, the boundary conditions imply $\pback{\ell \into
\emF}\hateq 0$ (equation (\ref{pbackf})).  Hence, it follows
immediately that $\Phi_{(\ell)}$ is constant on the horizon. We can
regard this result as the `electromagnetic part' of the zeroth law of
isolated horizon mechanics.

We will see in the next section that these zeroth laws play a key role
in making the gravitational and the electromagnetic action principles
well-defined in presence of isolated horizons.  As with surface gravity
$\kappa_\ls$, the functional dependence of $\Phi_{(\ell)}$ on the
horizon parameters varies with the choice of $\ell \in [\l]$.  We will
see in Section \ref{s5} that the Hamiltonian framework constrains these
dependencies in an interesting fashion.

\subsection{Other definitions and remarks}
\label{s2.C}

In this sub-section, we introduce two new definitions which are
important to the general framework of isolated horizons.

The first is concerned with rotation.  From one's experience with the
Newman-Penrose framework, one expects the gravitational contribution
to angular momentum to be coded in the imaginary part of $\Psi_2$.
This expectation will be shown to be correct in
\cite{abl2}. Therefore, in the Einstein-Maxwell theory, we introduce
the following definition:

\medskip\noindent\textbf{Definition 4}: A weakly isolated horizon
$(\Delta, [\l])$ will be said to be \textit{non-rotating} if
$\ImPt{\Psi_2}$ vanishes on $\Delta$. \\

\medskip\noindent If $(\Delta_{\rm K}, [\xi])$ is a Killing horizon
and $\xi$ is a \textit{hypersurface orthogonal}, time-like vector field
near $\Delta_{\rm K}$, on physical grounds one would expect the horizon
to be non-rotating. Is this expectation compatible with our definition?
The answer is in the affirmative. For, in this case, one can show that
$B_{ab}:=C_{acbd} \xi^c \xi^d$ vanishes in the region where $\xi^a$ is
time-like.  Hence, by continuity, it also vanishes on $\Delta$ forcing
$\ImPt{\Psi_2}$ to vanish there. Similarly, if the space-time admits a
\textit{hypersurface orthogonal}, rotational Killing field $\varphi^a$
in a neighborhood of $\Delta$ $\ImPt{\Psi_2}$ again vanishes on
$\Delta$. The definition is again compatible with one's intuition that
the horizon should be non-rotating in this case. In this paper, while
we allow for presence of rotation in the first four sections, we will
restrict ourselves to non-rotating horizons in the proof of the first
law in Section \ref{s5}.

Finally, for completeness, let us introduce a stronger notion of
`isolation' by strengthening the boundary conditions of Definition 2.
\medskip

\noindent\textbf{Definition 5}: A weakly isolated horizon $(\Delta, [\l
])$ is said to be \textit{isolated} if
\begin{equation}
\label {si} [{\cal L}_{\l}, \D] V \= 0
\end{equation}
for all vector fields $V$ tangential to $\Delta$ and all $\l \in [\l ]$
\medskip

As before, if any one $\l$ satisfies this condition, so do all $\l \in
[\l ]$.  However, unlike (\ref{i}), condition (\ref{si}) is a
\textit{genuine} restriction in the sense that it can not always be
met by a judicious choice of null normals.  Generically it does
suffice to single out the equivalence class $[\l ]$ uniquely
\cite{abl1}.  In particular, in the Kerr family, the only $[\l]$ which
satisfies (\ref{si}) is the one containing constant multiples of the
globally defined Killing field which is orthogonal to the
horizon. Every Killing horizon is of course an isolated horizon. Thus,
even though (\ref{si}) is a stronger condition than (\ref{i}), it is
still \textit{very} weak compared to conditions normally imposed. For
most physical applications, e.g., to numerical relativity, it is
appropriate to work with isolated horizons. For mechanics of isolated
horizons, however, we can ---and will--- work with the larger class of
weakly isolated horizons.

The following consequences of the boundary conditions defining weakly
isolated horizons, derived in this section, will play an important
role in the subsequent discussion:
\begin{description}
\item[a)] $\nabla_{\pback{a}} \ell_b \= \omega_a \ell^b$;
\item[b)] $\kappa_{(\ell)} \= \ell^a \omega_a$, the surface gravity
defined by the null normal $\ell^a$, is constant on $\Delta$;
\item[c)] There is a natural (area) 2-form $\twoeps$ on $\Delta$ satisfying
$\Lie_{\ell}\, \twoeps \= 0$ and $\twoeps_{ab} \ell^b \= 0$.
\item[d)] The electromagnetic potential $\emA$ is chosen to satisfy
$\Lie_{\ell}\emA \= 0$ and in this gauge the scalar potential
$\Phi_{(\ell)} := - \ell^a\emA_a$ is constant on $\Delta$; and
\item[e)] The electromagnetic field satisfies $\pback{\ell^a \emF_{ab}}
\= 0$; and $\pback{\ell^a \dual\emF_{ab}} \= 0$.
\end{description}

\section{Action}
\label{s3}

In this paper we use the first order formulation of general relativity
in terms of tetrads and connections.  Since tetrads are essential to
incorporate spinorial matter, it is natural to base the framework on
tetrads from the beginning.  The use of a first order formalism, on the
other hand, is motivated primarily by mathematical simplicity.  In the
first order framework, the action and the Hamiltonians can be expressed
entirely in terms of differential forms which significantly simplify
the variational calculations.  The previous paper \cite{abf2} which
dealt with undistorted horizons used spinors and self-dual connections,
while here we choose to use orthonormal tetrads and real, Lorentz
connections.  For analyzing mechanics of isolated horizons, there are
two advantages to this.  First, the Hamiltonian and symplectic
structure are now manifestly real which simplifies evaluation of the
boundary terms at the horizon.  Second, the analysis can now be
extended to other space-time dimensions in a straightforward manner.
However, these simplifications, come with a price.  Since, at present,
the self dual variables appear to be indispensable for non-perturbative
quantization, the results obtained here will have to be re-expressed in
terms of self-dual variables in order to extend the analysis \cite{abk}
of the quantum horizon geometry and black hole entropy to  include
rotation.

\subsection{Preliminaries}
\label{s3.A}

Let us begin with the first order action for Einstein-Maxwell theory on
a 4-dimensional manifold $\man$ which is topologically $M\times\Rbar$,
where $M$ is an oriented Riemannian 3-manifold \textit{without
boundary} (the complement of a compact set of) which is diffeomorphic
to (the complement of a compact set of) $\Rbar^3$.  Thus, topological
complications of $M$, if any, are confined to a compact set.  In this
subsection we shall only give the relevant formulae. For details, see
e.g. \cite{aws}.  Our basic fields will consist of a triplet
$(e_a^I,A_{aI}{}^J,\emA_a)$ defined on $\man$ where $e_a^I$ denotes a
co-tetrad, $A_{aI}{}^{J}$ the gravitational (Lorentz) connection and
$\emA_{a}$ the electro-magnetic connection. Here, lower case latin
letters refer to the tangent space of $\man$ while the upper case
letters $I,J$ etc. refer to an internal four dimensional vector space
$V$ with a fixed metric $\eta_{IJ}$ of signature $(-+++)$. The
co-tetrad $e_a^I$ is an isomorphism between the tangent space
$T_p(\man)$ at any point $p$ and the internal space $V$. Using it, we
define a metric on $\man$ by $g_{ab}:= e_a^I e_b^J \eta_{IJ}$ which
also has signature $(-+++)$. The Lorentz connection $A_{aI}{}^{J}$ acts
only on internal indices and defines a derivative operator
\[
D_a k_I :=\partial_a k_I + {A_{aI}}^J k_J \, ,
\]
where $\partial$ is a fiducial derivative operator which, as usual,
will be chosen to be flat and torsion free.  Finally, $\emA_a$ is the
$U(1)$ electromagnetic connection 1-form on $\man$.  (As noted in
Section \ref{s2.D}, we will assume that the magnetic charge is zero.)
All fields will be assumed to be smooth and satisfy the standard
asymptotic conditions at infinity.

The 2-forms $\Sigma^{IJ}$
\[\Sigma_{IJ}:=\half\epsilon_{IJKL} e^K \wedge e^L\]
constructed from the co-tetrads will play an important role throughout
our calculations.  In particular, the action for an asymptotically flat
space-time (with no internal boundary) is given by (see e.g.
\cite{aws})
\begin{equation}\label{action}
S (e, A, \emA) =
    \frac{-1}{16\pi G} \int_{\man} \Sigma^{IJ}\wedge F_{IJ}
    +\frac{1}{16\pi G} \int_{\tau_\infty} \Sigma^{IJ} \wedge A_{IJ}
    -\frac{1}{8\pi}\int_{\man} \emF\wedge \dual\emF \, .
\end{equation}
Here $F$ and $\emF$ are the curvatures of the gravitational and
electromagnetic connections $A$ and $\emA$ respectively:
\[
F_{I}{}^{J} = dA_{I}{}^{J} + A_{I}{}^{K} \wedge A_{K}{}^{J} \qquad
\textrm{and} \qquad \emF = d\emA  \, ,
\]
$\dual\emF_{ab} = \textstyle{1\over 2} \epsilon_{ab}{}^{cd}\,
\emF_{cd}$ is the dual of $\emF$ defined using $e_{a}{}^I$, and
$\tau_\infty$ is the time-like cylinder at infinity.  The boundary term
at $\tau_\infty$ ensures the differentiability of the action.

Let us briefly examine the equations of motion arising from the action.
Varying the action with respect to the connection, one obtains
\[ D \Sigma =0. \]
This condition implies that the connection $D$ defined by $A$ has the
same action on internal indices as the unique connection $\nabla$
compatible with the co-tetrad, i.e., satisfying $\nabla_a e_b^I =0$.
When this equation of motion is satisfied, the curvature $F$ is related
to the Riemann curvature $R$ of $\nabla$ by
\[{F_{ab}}^{IJ}= {R_{ab}}^{cd}e_c^I e_d^J\, . \]
Varying the action with respect to $e_a^I$ and taking into account the
above relation between curvatures, one obtains Einstein's equations
\[ G_{ab} = 8\pi G T_{ab} \equiv  {2}{G}\left(\emF_{ac} \emF_{bd}
g^{cd} - \frac{1}{4} g_{ab} \emF_{cd} \emF^{cd}\right) \]
where $G_{ab}$ is the Einstein tensor and $T_{ab}$ the electromagnetic
stress energy tensor.  Finally, variation with respect to the
electromagnetic connection, $\emA$, yields Maxwell's equation
\[d \dual{\emF} = 0 \, .\]

\subsection{Internal boundary $\Delta$}
\label{s3.B}

Let us now consider the variational principle for asymptotically flat
histories which admit a weakly isolated horizon $\Delta$ as their
internal boundary.  The manifold $\man$ under consideration has an
internal boundary $\Delta$, topologically $S^2\times \Rbar$.  As
before, $\man$ is topologically $M\times \Rbar$,  where $M$ is now an
oriented manifold with an internal, 2-sphere boundary, whose
topological complications are again confined to a compact region.
Space-time is bounded to the future and past by two (partial Cauchy)
surfaces $M^\pm$, extending to spatial infinity (see Figure 1).

\begin{figure} \label{f1}
  \begin{center}
  \includegraphics[height=6cm]{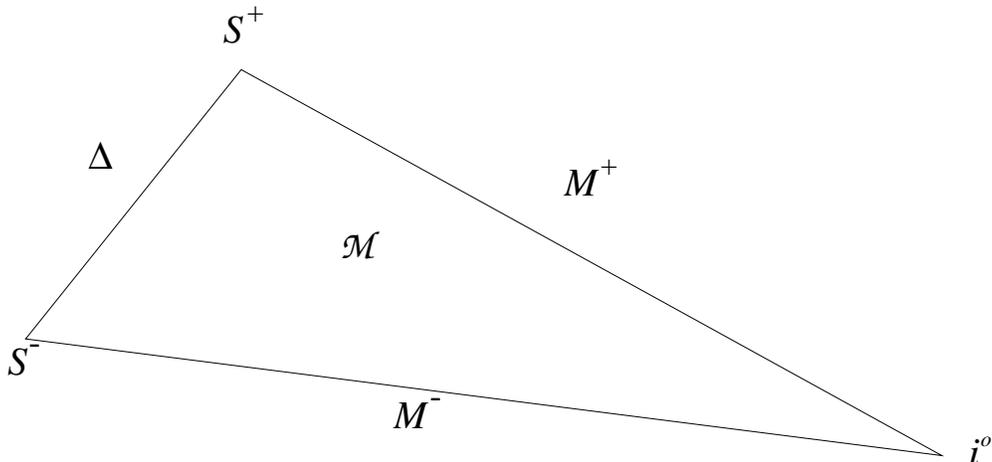}
  \caption{The region of space-time $\man$ under consideration has an
  internal boundary $\Delta$ and is bounded by two partial Cauchy
  surfaces $M^{\pm}$ which intersect $\Delta$ in the $2$-spheres
  $S^{\pm}$ and extend to spatial infinity $i^{o}$. }
  \end{center}
\end{figure}

Following Definition 2 of weakly isolated horizons, we will equip
$\Delta$ with a fixed equivalence class of vector fields $[\l ]$ which
are transversal to its 2-sphere cross-sections (where, as before, $\l
\sim \l^\prime$ if and only if $\l \= c \l^\prime$ for a constant $c$).
It is also convenient to fix an \textit{internal} null tetrad $(\ell^I,
n^I, m^I, \bar{m}^I)$ on $\Delta$, each element of which is annihilated
by the fiducial, flat internal connection $\partial$.

The permissible histories consist of smooth triplets $(e, A, \emA)$ on
$\man$ satisfying boundary conditions at infinity and on $\Delta$. The
boundary conditions at infinity are, as before, the standard ones which
ensure asymptotic flatness.  Since the asymptotic behavior and boundary
integrals at infinity play only a secondary role in our analysis, we
shall not spell out the precise fall-off requirements.  At $\Delta$,
the histories are subject to three conditions: i) the tetrads $e$
should be such that the vector field $\l^a:= \l^I e^a_I$ defined by
each history belongs to the equivalence class $[\l ]$ fixed on
$\Delta$; ii) the tetrad $e$ and the gravitational connection $A$
should be such that $(\Delta, [\l ])$ is a weakly isolated horizon for
the history; and, iii) the electromagnetic potential $\emA$ is in a
gauge adapted to the horizon, i.e., ${\cal L}_\l \emA \= 0$.

\textsl{Remark}: In space-time, we have the freedom to perform a local,
internal Lorentz rotation on the tetrad $e^a_I$ (and the gravitational
connection $A_{aI}{}^J$).  All these tetrads define the same Lorentzian
metric $g_{ab}$.  Since $\l^a$ is required to be a null normal to
$\Delta$, the permissible gauge rotations are reduced on $\Delta$ to
the sub-group $(\Rbar^+ \times E^2)_{\rm loc}$ of local null rotations
preserving the null direction field $\ell$. (Here $\Rbar^+$ is the
group of rescalings of $\l^a, n^a$ which leaves $m^a$ fixed and $E^2$
is the 3-dimensional Euclidean group consisting of rotations in the
$\l$-$m$, $\l$-$\bar{m}$ and $m$-$\bar m$ planes.) Condition i) above
--- dictated by the existence of a preferred equivalence class $[\l ]$
in Definition 2 --- further reduces the internal gauge freedom to
$\Rbar^+ \times (E^2_{\rm loc})$, i.e., reduces the group
$\Rbar^+_{\mathrm{loc}}$ of local $\l$-$n$ rescalings to the group
$\Rbar^+$ of global rescalings.  Thus, while any one space-time $(\man,
g_{ab})$, still defines infinitely many histories due to the freedom of
tetrad-rotations, this freedom is somewhat reduced at $\Delta$ because
of the structure fixed by the boundary
conditions.%
\footnote{Nonetheless, from a space-time perspective, the multiplicity
of histories can still be rather surprising. For example, if $g_{ab}$
is the Schwarzschild metric with mass $M>0$, there is a history in
which the surface gravity $\kappa_{\ls}$ is positive and another in
which it is zero. This redundancy can be eliminated by working with
isolated, rather than weakly isolated horizons.}

Given any tetrad $e^a_I$, the internal null vectors $(\l^I, n^I, m^I,
\bar{m}^I)$ fixed on $\Delta$ trivially  provide a null tetrad $(\l^a,
n^a, m^a, \bar{m}^a)$. In terms of these vectors, we can express
$\pback{\Sigma}^{IJ}$ as:
\begin{equation}\label{sigdelta}
\pback{\Sigma}^{IJ} \hateq 2\ell^{[I}n^{J]}\, \twoeps + 2 n\wedge (i m
\ell^{[I}\bar{m}^{J]}  -  i \bar{m}  \ell^{[I} m^{J]}),
\end{equation}
where, as before, $\twoeps = i m\wedge \bar{m}$ is the pull-back to
$\Delta$ of the natural alternating tensor on the 2-sphere ${\cal S}$
of integral curves of $\l^a$ associated with the given history. The
weak isolation of $(\Delta, [\l ])$ restricts the form of the
connection $A$ at $\Delta$.  To see this, recall that one of the
equations of motion requires the connection $D$ defined by $A$ to have
the same action on internal indices as $\nabla$. Hence,
$\nabla_{\!\pback{a}} \ell_I \= \partial_a\l_I + A_{{\pback{a}}I}{}^J
\ell_J \= A_{{\pback{a}}I}{}^J \ell_J$ where, in the second step we
have used the fact that the flat derivative operator, $\partial$, has
been chosen to annihilate the internal tetrad on $\Delta$. Since
$\nabla_a e_b^I = 0$ by definition of $\nabla$, and
$\nabla_{\!\pback{a}} \ell^b \= \omega_a \ell^b$ (see \ref{omegadefn})
it follows that $\pback{A}_{IJ} \ell^J \= \omega \ell_I$.  Hence, on
$\Delta$, $\pback{A}$ has the form:
\begin{equation}\label{conndelta}
\pback{A}_{IJ} \hateq -2l_{[I}n_{J]}\omega + C_{IJ},
\end{equation}
where the 1-form $C_{IJ}$ satisfies $C_{IJ}\l^J \= 0$.

With this background material at hand, we are now ready to consider
variations of the action (\ref{action}) in the presence of an inner
boundary representing a weakly isolated horizon.  A key question is
whether a new surface term at the horizon is necessary to make the
variational principle well-defined. We will show that, thanks to the
zeroth law, such a term is not needed.

A simple calculation yields:
\begin{equation} \label{var}
    \delta S(e,A,\emA)  =
    \int_{\man} \mbox{Equations of Motion} \cdot \delta \phi
    -\frac{1}{16\pi G}\int_{\Delta} \Sigma^{IJ}\wedge \delta A_{IJ}
    -\frac{1}{4\pi} \int_{\Delta} \delta \emA \wedge \dual \emF.
\end{equation}
where, in the first term, $\phi$ stands for the basic fields $(e, A,
\emA)$ in the action.  Note that, as in the case without an internal
boundary, the variation of the boundary term at infinity precisely
cancels the contribution arising from the variation of the bulk terms.

In order to show that the action principle is viable, it is necessary
to show that the terms at the horizon vanish due to the boundary
conditions imposed there.  Let us begin with the gravitational term.
Using (\ref{sigdelta}) and (\ref{conndelta}) it can be re-expressed as
\begin{equation}\label{gravdelta}
-\frac{1}{8\pi G} \int_{\Delta} \delta \omega \wedge \twoeps.
\end{equation}
Since $\twoeps$ is the pull-back to $\Delta$ of the alternating tensor
on the 2-sphere ${\cal S}$ of integral curves of $\l$, it follows that
${\cal L}_{\l} \twoeps \= 0$.  Furthermore, the weak isolation of the
horizon ensures ${\cal L}_{\l} \omega \= 0$ and, since the null normal
$\l^a$ defined by \textit{any} tetrad belongs to the fixed equivalence
class $[\l ]$ at the horizon, we have $\delta \l \= c_{\delta} \l$ for
some constant $c_\delta$.  These two facts imply ${\cal L}_{\l} \delta
\omega \= 0$.  Thus the entire integrand is Lie dragged by $\ell$. In
the variational principle, however, all fields are fixed on the initial
and final hypersurfaces, say $M^\pm$. In particular, $\delta \omega$
necessarily vanishes on the initial and final cross sections of the
horizon. Therefore, the integrand in (\ref{gravdelta}) vanishes on the
initial and final cross sections \textit{and} is Lie dragged by $\ell$.
This immediately implies (\ref{gravdelta}) is zero.

Let us now consider the electromagnetic term.  Since every $\emA$ is in
a gauge adapted to the isolated horizon, ${\cal L}_{\l} \emA \= 0$.
Furthermore, $\delta\l^a \= c_{\delta} \l^a$, so we conclude ${\cal
L}_{\l} \delta\emA \= 0$.  Next, (\ref{lief}) ensures ${\cal L}_{\l}
\dual\emF \=0$.  Thus, the integrand of the electromagnetic surface
term is Lie dragged by $\l^a$.  An identical argument to the one
presented above implies that the electromagnetic surface term in
(\ref{var}) also vanishes. Therefore, the variation of the action
(\ref{action}) continues to yield Einstein-Maxwell equations in spite
of the presence of an inner boundary representing a weakly isolated
horizon.

It is instructive to re-examine the key step in the above argument.
Suppose we only had a non-expanding horizon.  Then, the gravitational
surface term could still be reduced to (\ref{gravdelta}), and $\twoeps$
and $\pback{\dual\emF}$ would still be Lie-dragged by $\ell$.  However,
in this case, we could not argue that $\omega$ and $\emA$ are also
Lie-dragged.  As we saw in Sections \ref{s2.B} and \ref{s2.D}, these
conditions are equivalent, respectively, to the constancy of the
surface gravity $\kappa_{\ls}$ and the electromagnetic potential
$\Phi_{\ls}$ on $\Delta$. In this sense, given a non-expanding horizon
as the inner boundary, the gravitational and electromagnetic zeroth
laws are the \textit{necessary and sufficient conditions} one must
impose for the viability of the standard, first order, tetrad action
principle.

\textsl{Remark}: Note that (\ref{action}) is not the unique viable
action for the problem: as usual, there is freedom to add suitable
boundary terms without affecting the viability.  Specifically, we are
free to add any horizon boundary term which is composed entirely of
fields which are Lie dragged by $\ell$, for example the intrinsic
horizon metric $q_{ab}$ and fields $\omega$, $\twoeps$, and
$\pback\emA$.  Then, due to the argument given above, the new action
would also be viable.  However, as usual, this freedom will not affect
the definition of the symplectic structure which underlies the
Hamiltonian treatment of the next section.

\section{Covariant Phase space}
\label{s4}

Let us now construct the phase space of space-times containing weakly
isolated horizons.  In the next section, we will use this framework to
construct Hamiltonians generating suitable time translations and
define the energy of an isolated horizon.  In \cite{abf2}, the phase
space was constructed by performing a Legendre transform of the
action.  This procedure leads to a `canonical' framework in which the
phase space consists of configuration and momentum variables defined
on a spatial hypersurface.  With the self-dual connections used in
\cite{abf2}, the gravitational configuration variable turns out to be
a connection and its conjugate momentum, a 2-form so that the
Hamiltonian description can again be given in terms of forms.  With
the full Lorentz connections now under consideration, the situation
turns out to be more complicated.  Specifically, one encounters
certain second class constraints and, when these are solved, one ends
up with the same canonical phase space that one would have obtained
through a second order formalism.  In the Hamiltonian framework, then,
the simplicity we encountered in Section \ref{s3} is lost.  More
specifically, constraint functions and Hamiltonians now contain terms
involving \textit{second} derivatives of the basic canonical variables
which make variations rather complicated.  (For details, see chapters
3 and 4 in \cite{aws}.)  Therefore, in this section we will not use a
Legendre transform.  Instead, we will construct the `covariant phase
space' from the space of solutions to field equations (see, e.g.,
\cite{abr}).  As in Section \ref{s3}, all expressions will now involve
only the basic form-fields and their exterior derivatives and
variational calculations will continue to be simple.

To specify the phase space, let us begin as in Section \ref{s3} by
fixing a manifold $\man$ with an internal boundary $\Delta$ (see figure
1).  As before, we will equip $\Delta$ with an equivalence class $[\l
]$ of vector fields transverse to its 2-sphere cross-sections.  To
evaluate the symplectic structure and Hamiltonians, we will often use a
partial Cauchy surface $M$ in the interior of $\man$ which intersects
$\Delta$ in a 2-sphere $S$. Points of the covariant phase space
$\Gamma$ will consist of histories considered in Section \ref{s3}
\textit{which satisfy field equations}.  More explicitly, $\Gamma$
consists of asymptotically flat solutions $(e,A,\emA)$ to the field
equations on $\man$ such that i) the vector field $\l^a:= \l^I e^a_I$
belongs to the equivalence class $[\l ]$ fixed on $\Delta$, ii) in each
solution, $(\Delta, [\l ])$ is a weakly isolated horizon; and, iii) the
electromagnetic potential $\emA$ is in a gauge adapted to the horizon,
i.e., ${\cal L}_\l \emA \= 0$.

Our next task is to use the action (\ref{action}) to define the
symplectic structure $\Omega$ on $\Gamma$.  It is convenient to make a
brief detour and first introduce two new fields which can be regarded
as `potentials' for the surface gravity $\kappa_{\ls}$ and the electric
potential $\Phi_{\ls}$.  Given any point $(e,A,\emA)$ in the phase
space $\Gamma$, let us define scalar fields $\psi$ and $\chi$ on
$\Delta$ as follows: \\
i) ${\cal L}_{\l} \psi \hateq (\ell \cdot \omega) \hateq
    \kappa_{\ls}$ and ${\cal L}_{\l} \chi \= (\ell \cdot \emA) \=
    -\Phi_{\ls}$; and \\
ii) $\psi$ and $\chi$ vanish on $S^-$, the intersection of
    $M^-$ with $\Delta$.%
\footnote{Condition ii) serves only to fix the freedom to add constants
to $\psi$ and $\chi$.  One could envisage replacing it by a different
condition.  Our results will be insensitive to this choice.}
\\
Note that these conditions associate with each point of $\Gamma$ a
unique pair $(\psi, \chi)$ on $\Delta$ and in the `extremal' case
$\kappa_{\ls} = 0$, $\psi$ vanishes identically.

We wish to use the standard procedure \cite{abr} involving second
variations of the action to define the symplectic structure.%
\footnote{Actually this procedure provides a pre-symplectic structure,
i.e., a closed 2-form on the phase-space which, however, is generally
degenerate.  The vectors in its kernel represent infinitesimal `gauge
transformations'.  The physical phase space is obtained by quotienting
the space of solutions by gauge transformations and inherits a true
symplectic structure from the pre-symplectic structure on the space of
solutions.  The 2-form $\Omega$ introduced below is indeed degenerate.
However, for simplicity, we will abuse the notation somewhat and refer
to $\Omega$ as the symplectic structure.}
Let us recall the main steps of this procedure. One first constructs
the symplectic current $J$: Given a point $\gamma$ in the phase space
$\Gamma$ and two tangent vectors $\delta_1$ and $\delta_2$ at that
point, $J$ provides a closed 3-form $J(\gamma; \delta_1,\delta_2)$ on
$\man$. Integrating $dJ$ over the part $\tilde{\man}$ of space-time
under consideration, one obtains
\[
 0= \int_{\tilde{\man}} d\, J(\gamma; \delta_1,\delta_2) =
\oint_{\partial{\tilde{\man}}} J.
\]
Now, if there is no internal boundary, one can choose $\tilde{\man}$ to
be a region bounded by any two Cauchy surfaces $M_1$ and $M_2$ so that
the boundary is given by $\partial\tilde{\man} = M_1 \cup M_2 \cup
\tau_\infty$, where $\tau_\infty$ is the time-like `cylinder at
infinity'.  In simple cases, the asymptotic conditions ensure that the
integral $\int_{\tau_\infty} J(\gamma; \delta_1,\delta_2) $ vanishes.
Then, taking orientations into account, it follows that $\int_M\,
J(\gamma; \delta_1,\delta_2)$ is independent of the choice of Cauchy
surface $M$.  One then sets the symplectic structure to be
\[\Omega|_\gamma\,(\delta_1, \delta_2) = \int_M J(\gamma;
\delta_1,\delta_2) \, . \]

In our case, the second variation of the action (\ref{action}) yields
the following symplectic current:
\begin{equation}\label{symcur}
J(\gamma; \delta_1,\delta_2)=
    \frac{-1}{16\pi G}[ \delta_1 \Sigma^{IJ}\wedge \delta_2 A_{IJ}
        -\delta_2 \Sigma^{IJ} \wedge \delta_1 A_{IJ} ]
    -\frac{1}{4\pi } [\delta_1 \dual \emF \wedge \delta_2 \emA
        -\delta_2 \dual \emF \wedge \delta_1 \emA ].
\end{equation}
Using the fact that the fields $\gamma \equiv (e,A, \emA)$ satisfy the
field equations and the tangent vectors $\delta_1, \delta_2$ satisfy
the linearized equations off $\gamma$, one can directly verify that
$J(\gamma; \delta_1, \delta_2)$ is in fact closed as guaranteed by the
general procedure involving second variations.  It is now natural to
choose $\tilde{\man}$ to be a part of our space-time $\man$ bounded by
two partial Cauchy surfaces $M_1, M_2 $, the time-like cylinder
$\tau_\infty$ and a part $\tilde\Delta$ of the isolated horizon bounded
by $M_1$ and $M_2$.  Again, the asymptotic conditions ensure that the
integral of $J$ over $\tau_\infty$ vanishes.  Hence,
\[ (\int_{M_1} + \int_{M_2} + \int_{\tilde\Delta}) J(\gamma;
\delta_1,\delta_2) = 0. \]
However, this does not immediately provide us the conserved symplectic
structure because the integral of $J$ over $\tilde\Delta$ does not
vanish in general.  Since the isolation of the horizon implies that
there are no fluxes of physical quantities across $\Delta$, one might
expect that, although non-zero, the integral over $\tilde\Delta$ would
be `controllable'.  This is indeed the case. Using the forms
(\ref{sigdelta}) and (\ref{conndelta}) of $\pback{\Sigma}$ and
$\pback{A}$ on $\Delta$ and the definitions of the potentials $\psi$
and $\chi$, it is easy to verify that the pull-back of the symplectic
current to $\Delta$ is itself exact:
\[ \pback{J} (\gamma; \delta_1, \delta_2) \= d j(\gamma, \delta_1,
\delta_2)\]
where the 2-form $j$ on $\Delta$ is given by:
\[ j(\gamma, \delta_1, \delta_2) = \frac{1}{8\pi G} [\delta_1 \psi\,
\delta_2(\twoeps) - \delta_2 \psi \delta_1 (\twoeps)\, ] +
\frac{1}{4\pi}[ \delta_1\chi \delta_2\, \dual\emF - \delta_2\chi
\delta_1\,\dual \emF ]. \]
Hence, if $M_1$ and $M_2$ intersect $\tilde\Delta$ in 2-spheres $S_1$
and $S_2$ respectively, we have:
\[ \int_{\tilde\Delta} J(\gamma; \delta_1, \delta_2)
= - (\int_{S_1} + \int_{S_2}) j(\gamma; \delta_1, \delta_2) .\]
The negative sign appearing in the above expression is due to the
choice of orientation of $S_{\Delta}$, which is induced from $M$ rather
than from $\Delta$. Using these results we can define the symplectic
structure as:
\begin{eqnarray}\label{sym}
   \lefteqn{ \Omega|_{\gamma} \, (\delta_1 ,\delta_2 )= } \nonumber\\
    &&\frac{-1}{16\pi G}\int_{M} [\delta_1 \Sigma^{IJ}\wedge \delta_2 A_{IJ}
        -\delta_2 \Sigma^{IJ} \wedge \delta_1 A_{IJ}] \,
    +\frac{1}{8\pi G} \oint_S [\delta_1 (\twoeps)\, \delta_2\psi -
        \delta_2 (\twoeps)\, \delta_1\psi] \nonumber \\
    && - \frac{1}{4\pi} \int_{M} [\delta_1 \dual \emF \wedge \delta_2 \emA
        -\delta_2 \dual \emF \wedge \delta_1 \emA] \,
    + \frac{1}{4\pi} \oint_S [\delta_1 \dual\emF\, \delta_2\chi -
    \delta_2 \dual\emF\, \delta_1 \chi]
\end{eqnarray}
Again, using field equations one can directly verify that the right
side of (\ref{sym}) is independent of the choice of the partial Cauchy
surface; the symplectic structure is `conserved'.  We will use
$(\Gamma, \Omega)$ as our covariant phase space.

Note that, even though the action did not contain a surface term at the
horizon, the symplectic structure does.  So, the overall situation is
the same  as in the undistorted, non-rotating case considered in
\cite{abf2}.  Finally, our discussion of the action principle and our
construction of the covariant phase space is applicable to \textit{all}
weakly isolated horizons $\Delta$; nowhere did we have to restrict
ourselves to the non-rotating case.

\section{Hamiltonian evolution and the first law}
\label{s5}

To discuss the first law, we must first define horizon energy, which in
turn requires a time evolution field $t^a$ on $\man$.  Given a vector
field $t^a$ satisfying appropriate boundary conditions, $\delta_t :=
(\Lie_t e, \Lie_t A, \Lie_t \emA)$ satisfies the linearized equations
for any $\gamma := (e,A,\emA)$ in $\Gamma$ and thus defines a vector
field on $\Gamma$.%
\footnote{In the Lie-derivatives, the internal indices are treated as
scalars; thus $\Lie_t e_a^I = t^b\partial_b e_a^I + e_b^I \partial_a
t^b$. To make $\delta_t$ a well-defined vector field on $\Gamma$, we
now exclude $M^\pm$ from $\man$ and let $\man$ and $\Delta$ be without
future and past boundaries.  Whenever needed, these boundaries, $M^\pm$
and $S^\pm$, can be added by taking the obvious closure of $\man$.}
This $\delta_t$ can be interpreted as the infinitesimal generator of
time evolution on the covariant phase space.  It is then natural to ask
if this vector field is a phase space symmetry, i.e., if
$\Lie_{\delta_t} \Omega$ vanishes everywhere on $\Gamma$.  The
necessary and sufficient condition for this to happen is that there
exist a function $H_t$ ---the Hamiltonian generating the
$t$-evolution--- such that
\begin{equation}\label{hamilton}
\delta H_t =\Omega(\delta, \delta_t)
\end{equation}
\textit{for all} vector fields $\delta$ to $\Gamma$.  On general
grounds, one expects $H_t$ to contain a surface term $E_{{\rm
ADM}}^{t}$ at infinity representing the total (i.e. ADM) energy, and a
surface term $E_{\Delta}^{t}$ at the horizon which can be interpreted
as the horizon energy, both tied to the evolution field $t^a$.

A key question then is to specify the appropriate boundary conditions
on $t^a$.  It is clear that, at infinity, $t^a$ should be an asymptotic
time-translation, i.e., should approach a time-translation Killing
field of the flat metric used to specify the boundary conditions.  At
the horizon, on the other hand, the metric is \textit{not} universal
and the space-time defined by a generic point $\gamma$ of the covariant
phase space does not admit \textit{any} Killing field near $\Delta$.
Therefore, specification of the boundary conditions at $\Delta$ is not
as straightforward as that at infinity. It is for this reason that we
now assume that $(\Delta, [\l ])$ is a \textit{non-rotating}, weakly
isolated horizon for all points $\gamma \equiv (e,A,\emA)$ of the phase
space.  The problem of specifying the appropriate boundary conditions
on $t^{a}$ in the rotating case is more complicated . However, it has
been addressed successfully and will be discussed in \cite{abl2}.

Recall that the internal boundary $\Delta$ of $\man$ is equipped with a
\textit{specific} equivalence class $[\l ]$ of vector fields.  As
discussed in Section \ref{s2}, these $\l$ are the isolated horizon
analogs of constant multiples of Killing fields on the Killing horizons
in static space-times.  Therefore, in the \textit{non-rotating} case,
it is natural to demand that, on $\Delta$, the evolution vector field
$t^a$ should belong to the equivalence class $[\l ]$.  This
automatically ensures that $(\Lie_{t}e, \Lie_{t} A, \Lie_{t}\emA)$
satisfy the appropriate boundary conditions to define a tangent vector
at each point of the phase space $\Gamma$. However, unlike at infinity,
the geometry at the horizon is \textit{not} fixed once and for all.
Therefore, it is natural to allow the precise value of the evolution
vector field $t^a$ on $\Delta$ to vary from one point of the phase
space to another.  In more familiar terms, this corresponds to allowing
the (boundary values of) lapse and shift fields to depend on dynamical
fields $(e,A,\emA)$ themselves, a procedure routinely used in numerical
relativity and gauge-fixed calculations in canonical gravity. Following
the current terminology in numerical relativity, we will refer to such
$t^a$ as \textit{live} evolution vector fields. The use of live fields
turns out to be necessary to ensure that $\delta_t$ is a phase space
symmetry, i.e., yields a Hamiltonian evolution on $(\Gamma, \Omega)$.

Let us fix a live $t^a$ whose restriction to the horizon belongs to the
equivalence class $[\l]$ at all points of the phase space.  To analyze
if $\delta_t$ is a Hamiltonian vector field, it is simplest to compute
the 1-form $X_t$ on $\Gamma$ defined by
\begin{equation}\label{defx}
X_t (\delta)= \Omega(\delta, \delta_{t}).
\end{equation}
Now $\delta_t$ is Hamiltonian --- i.e., $\Lie_{\delta_{t}}\, \Omega =
0$ on $\Gamma$ --- if and only if $X_t$ is closed, i.e.,
\[\dd X_t =0\]
where $\dd$ denotes the exterior derivative on (the infinite
dimensional) phase space $\Gamma$. If this is the case then, up to an
additive constant, the Hamiltonian is given by
\[\dd H_t = X_t.\]

To calculate the right side of (\ref{defx}), it is useful to note the
following identities from differential geometry:
\begin{equation}\label{liet}
\begin{array}{rcl@{\hspace{5mm}}rcl}
    \Lie_t A &=& t \into F + D(t\cdot A) &
    \Lie_t \Sigma &=& t\into D\Sigma + D(t\into \Sigma) -[(t\cdot A),\Sigma] \\
    [2mm]
    \Lie_t \emA &=& t\into \emF + d(t\cdot \emA)&
    \Lie_t \dual \emF &=& t\into(d\,\dual\emF)+d(t\into \dual \emF)
\end{array}
\end{equation}
Using these, the field equations satisfied by $(e,A, \emA)$ and the
linearized field equations for $\delta$, we obtain the required
expression of $X_t$:
\begin{eqnarray}\label{xt}
X_t(\delta) &:=& \Omega(\delta,\Lie_t)\\ &=& \frac{-1}{16\pi G}
\int_{\partial M} {\rm Tr}\left[(t\cdot A) \delta \Sigma
        -(t\into \Sigma) \wedge\delta A \right]
  - \frac{1}{4\pi}\int_{\partial M} (t\cdot \emA) \delta(\dual \emF)
        -(t\into \dual \emF)\wedge \delta\emA \, .
\end{eqnarray}
Note that the expression involves integrals \textit{only} over the
2-sphere boundaries $S_\infty$ and $S_\Delta$ of $M$, the partial
Cauchy surface used in the evaluation of the symplectic structure;
there is no volume term.

The integrals at infinity can be evaluated easily by making use of the
fall-off conditions.  As one would expect, the electromagnetic term
vanishes (because $\emA$ falls off at least as $1/r$ while $\emF$
falls off as $1/r^2$) while the gravitational term yields precisely
the ADM energy $E_{\rm ADM}^t$ associated with the asymptotic
time-translation defined by $t^a$.  At the horizon, we can use
equations (\ref{sigdelta}) and (\ref{conndelta}) to show that $t \into
\Sigma$ contracted on internal indices with $\delta A$ vanishes and
(\ref{pbackstarf}) implies $t \into \dual \emF =0$ leaving
\begin{eqnarray}
    X_t(\delta)\label{xtbc} &=&
    -\frac{1}{8\pi G} \int_{S_{\Delta}} (t \cdot\omega) \delta (\twoeps)
    - \frac{1}{4\pi}\int_{S_{\Delta}} (t \cdot\emA) \delta(\dual \emF)
    + \delta E_{\rm ADM}^t \nonumber\\
    &=& -\frac{1}{8\pi G}\kappa_{(t)} \delta a_{\Delta}
        - \Phi_{(t)} \delta Q_{\Delta}
        +\delta E_{\rm ADM}^t
\end{eqnarray}
where, in the last step, we have used the fact that both $t \cdot
\omega =\kappa_{(t)}$ and $t \cdot \emA = - \Phi_{(t)}$ are
constant on the horizon and the definition (\ref{charge}) of electric
charge.

The necessary and sufficient condition for the existence of a
Hamiltonian is that $X_t$ be closed. Clearly, this is equivalent to
\begin{equation}\label{iff}
\frac{1}{8\pi G}\,\dd \kappa_{(t)} \wwedge\dd a_{\Delta} +
\dd\Phi_{(t)} \wwedge \dd Q_{\Delta} = 0 \, ,
\end{equation}
where $\wwedge$ denotes the antisymmetric tensor product on $\Gamma$.
Now, (\ref{iff}) trivially implies that the surface gravity
$\kappa_{(t)}$ and the electric potential $\Phi_{(t)}$ at the horizon
defined by $t^a$ can depend \textit{only} upon the area and charge of
the horizon.  Other factors, such as the `shape' of the distorted
horizon, can not affect the values of $\kappa_{(t)}$ or $\Phi_{(t)}$.
Finally, (\ref{iff}) is the necessary and sufficient condition that
there exists a function $E_{\Delta}^{t}$ , also only of $a_{\Delta}$
and $Q_{\Delta}$ such that
\begin{equation} \label{1law}
\delta E_{\Delta}^{t} = \frac{1}{8\pi G}\,\kappa_{(t)} \delta a_{\Delta}
+ \Phi_{(t)} \delta Q_{\Delta} \, .
\end{equation}
Since $E_{\Delta}^{t}$ is a function \textit{only} of $a_{\Delta}$ and
$Q_{\Delta}$, it is a function of fields defined \textit{locally} at
the horizon.  As noted before, it is natural to interpret
$E_{\Delta}^{t}$ as the horizon energy defined by the time translation
$t^{a}$.  The total Hamiltonian is given by:
\be \label{ham} H_{t}  = E_{\rm ADM}^{t}  - E_{\Delta}^{t}. \ee

Let us summarize. Eq (\ref{1law}) is a necessary and sufficient
condition for the 1-form $X_{t}$ to be closed. Therefore, the vector
field $\delta_{t}$ on $\Gamma$ defined by the space-time evolution
field $t^{a}$ is Hamiltonian \textit{if and only if} the first law
(\ref{1law}) holds. Thus (\ref{1law}) is a restriction on the choice
of the \textit{live} vector field $t^a$: While any $t^a$ (which
preserves the boundary conditions) defines an evolution flow on the
phase space, it is only when
$$ \frac{1}{8\pi G}\,\kappa_{(t)} \dd a_{\Delta}
+ \Phi_{(t)} \dd Q_{\Delta} $$
is an exact 1-form on $\Gamma$ that this flow is Hamiltonian (i.e.,
preserves the symplectic structure). At first, this restriction seems
somewhat surprising because, in absence of internal boundaries,
\textit{every} vector field $t^a$ (which tends to a fixed Killing
field of the flat metric at infinity) defines a Hamiltonian
evolution. However, even in this context, there is no a priori reason
to expect this tight correspondence to hold if one allows general,
live vector fields $t^a$ whose boundary values at infinity can change
from one space-time to another. Finally, we will see in Section
\ref{s7} that every space-time belonging to the phase space $\Gamma$
admits an infinite family of vector fields $t^a$ for which $X_t$ is
closed.  Therefore, in particular, the first law does \textit{not}
restrict the `background' space-times (or the variations $\delta$) in
any way. Indeed, for any space-time in our phase space, there is an
infinite family of first laws, one associated with each permissible
$t^a$.

We will conclude this section with a few remarks.

i) \textsl{Form of $H_{t}$}: The Hamiltonian (\ref{ham}) contains only
surface terms.  This may seem surprising because, in the canonical
framework, the familiar Hamiltonian contains a volume integral
consisting of a linear combination of constraints. While the volume
term vanishes `on shell' and does not contribute to numerical value of
the canonical Hamiltonian on physical states, it is nonetheless crucial
for obtaining the correct evolution equations since derivatives of the
Hamiltonian transverse to the constraint surface are needed to
construct the Hamiltonian vector field.  The covariant phase space, by
contrast, consists only of solutions to the field equations whence the
issue of taking `off shell' derivatives never arises.  In
diffeomorphism invariant theories, the Hamiltonian
on the covariant phase space is always made of surface terms.%
\footnote{In particular, therefore, Hamiltonians generating
diffeomorphism which have support away from the boundaries vanish
identically. Unlike their counterparts on the canonical phase space,
the infinitesimal phase-space motions induced by such space-time vector
fields are in the kernel of the covariant symplectic structure.}
If space-time has several asymptotic regions, the boundary term in each
region defines the standard energy corresponding to that region.
Therefore, in the present case, it was natural to interpret
$E_{\Delta}^{t}$ as the  horizon-energy  defined by the $t^{a}$
evolution. Finally, we should emphasize that we used a covariant phase
space only for simplicity. The final results go through  (and, in fact,
were first obtained) in a canonical framework as well.

ii)\textsl{Comparisons}: As noted in the Introduction, all treatments
of the first law for non-rotating but possibly distorted horizons
available in the literature refer to static space-times.  The isolated
horizon framework, by contrast, does not refer to a Killing field at
all and thus allows a significantly larger class of physically
interesting situations.  On the other hand, since it relies on a
Hamiltonian framework, we cannot incorporate phenomenological matter
if it does not admit a phase space description.  Other treatments
based on Hamiltonian methods generally restrict themselves to static
space-times with a non-zero surface gravity.  This assumption is
essential there because those treatments use `bifurcate' surfaces in
an important way and these do not exist in the extremal static
solutions where the surface gravity vanishes.  In contrast, the
results of this section, do not refer to a bifurcate surface and go
through irrespective of whether $\kappa_{(t)}$ is non-zero or zero.
In a realistic collapse, the physical space-time is \textit{not}
expected to have the bifurcate surface.  The present analysis uses
only the portion of the physical space-time in which the horizon has
settled down with no further in-going radiation, rather than an
analytical continuation of the near horizon geometry used in certain
approaches.  Finally, in contrast to other treatments, we have an
infinite family of first laws, one for each evolution field $t^a$ for
which $\delta_t$ is a phase space symmetry.

iii)\textsl{Non-uniqueness of energy}: Each permissible, live $t^a$
defines a horizon energy $E^t_{\Delta}$. At first it seems surprising
that there is so much freedom in the notion of energy.  Let us compare
the situation at ${\cal I}^+$, which, like $\Delta$, is null. There, we
only have a 3-parameter freedom which, furthermore, can be eliminated
simply by fixing a rest frame. How does this difference arise? Recall
that energy is (the numerical value of) the generator of an
\textit{unit} time translation. At infinity, all 4-metrics in the phase
space approach the \textit{same} flat metric. Hence, we can simply fix
a unit time-translation Killing field $t^a_o$ of that flat metric near
infinity and use its restriction to ${\cal I}^+$ as the unit
time-translation for \textit{all} metrics in the phase space. By
contrast, there is no fixed 4-metric near $\Delta$ to which all the
metrics in our phase space approach. Hence, we do not have the analog
of $t^a_o$; only the equivalence class $[\l]$ is now common to all the
metrics. If, for a given metric $\tilde{g}_{ab}$ in our collection, we
select the time translation represented by a specific $\tilde{\l}^a$ in
$[\l]$, a priori we do not know which vector field $l^a$ in $[\l]$
would represent the `same' time-translation for another geometry
$g_{ab}$. One might imagine using the seemingly simplest strategy: just
fix a $\l_o^a$ in $[\l]$ and demand that $t^a$ approach that $\l_0^a$
\textit{for all} points $\gamma$ in the phase space. Unfortunately, the
strategy is not viable because such a $t^a$ fails
to define a Hamiltonian evolution in $\Gamma$.%
\footnote{As we saw above, a necessary condition for $\delta_t$ to be a
Hamiltonian vector field on $\Gamma$ is that $\kappa_{(t)}$ is a
function \textit{only} of $a_\Delta$ and $Q_\Delta$. Therefore, if we
can find a tangent vector $\delta$ in the phase space with $\delta
a_\Delta = \delta Q_\Delta =0$ but $\delta \kappa_{(t)} \not= 0$, $t^a$
can not define a Hamiltonian evolution. It is easy to find such a
tangent vector $\delta$ for this $t^a$.}
Finally, if we restrict ourselves to \textit{globally} static
space-times, we can overcome this difficulty by always working with the
Killing field which is unit at infinity. However, in absence of global
Killing fields, the behavior of the evolution vector field $t^a$ near
the horizon is unrelated to its behavior near infinity. Nonetheless, as
we shall show in Section \ref{s7}, if one has sufficient control on the
space of static solutions of the theory under consideration, it
\textit{is} possible to select a preferred energy function on the phase
space and use it as the mass of the isolated horizon. In the
Einstein-Maxwell case, all static solutions with horizons are
explicitly known whence the strategy is viable.

\section{Yang-Mills Field}
\label{s6}

In the previous three sections we restricted our attention to
Einstein-Maxwell theory. We will now indicate how Yang-Mills fields can
be included.  This section is divided into three parts.  In the first,
we discuss restrictions on the Yang-Mills fields due to the horizon
boundary conditions and introduce the notion of a `Yang-Mills gauge
adapted to the horizon'. In the second part, we discuss the action
principle and construct the covariant phase space for
Einstein-Yang-Mills theory. Using this formalism, in the third
subsection, we introduce a Hamiltonian generating time evolution and
extend the first law to the Yang-Mills case.

\subsection{Preliminaries}

We will restrict ourselves to compact gauge groups $G$ and Yang-Mills
connections defined on trivial bundles.  Since the bundle is trivial,
the connection gives rise to a smooth, globally defined Lie algebra
valued 1-form, $\ymA$. As usual, the Yang-Mills derivative operator
$\ymD$ will be defined as $\ymD\lambda = \partial\lambda +
[\ymA,\lambda]$, where $\partial$ is a flat Yang-Mills connection, and
the field strength, $\ymF$, via
\begin{equation}\label{ymfdef}
\ymF := d\ymA + \ymA \wedge \ymA.
\end{equation}
The stress energy tensor, $\ymT$, is given in terms of the field
strength as
\begin{equation}\label{ymt}
\ymT_{ab} = \frac{1}{4\pi}[{\ymF_{ac}}^i {{\ymF_b}^c}_i
    - \frac{1}{4}g_{ab} {\ymF_{cd}}^i{\ymF^{cd}}_i],
\end{equation}
where the label $i$ runs over the internal indices in the Lie algebra
of the group G.

Let us begin by examining how the isolated horizon boundary conditions
restrict the form of the field strength, $\ymF$, on $\Delta$. Since the
Yang-Mills stress energy tensor has the same form as the Maxwell one,
(\ref{emt}), the analysis is completely analogous to that of Section
\ref{s2.D}.  Therefore, we shall not include derivations of the
results, but instead highlight the differences.

Recall that, on a non-expanding horizon, $R_{ab}\l^a\l^b \= 0$ whence
$T_{ab}\l^a\l^b \= 0$.  This has several consequences for the
Yang-Mills field.  In particular, one concludes
\begin{equation}\label{pbackymf}
\pback{\ell^a \ymF_{ab}} \hateq 0 \qquad \mbox{and} \qquad
\pback{\ell^a  \dual{\ymF}_{ab}} \hateq 0.
\end{equation}
These two restrictions guarantee there is no flux of Yang-Mills field
across the horizon.  Making use of the specific form of the stress
energy tensor, we also conclude
\begin{equation}\label{phi20ym}
\Phi_{02} = \frac{1}{2} R_{ab}m^am^b \hateq 0 \qquad\qquad \mbox{and}
\qquad\qquad \Phi_{20}= \frac{1}{2} R_{ab}\bar{m}^a\bar{m}^b\hateq 0.
\end{equation}

Our next task is to define the Yang-Mills equivalents of the electric
and magnetic charges of the horizon.  Naively, one might consider
integrating $\ymF$ and $\dual\ymF$ over a 2-sphere cross section of
the horizon as was done in the Maxwell theory. However, these 2-forms
now have a free internal index and are only gauge covariant rather
than gauge invariant. Since there is no preferred internal basis at
the horizon, the integrals would fail to be well-defined. Therefore,
we must look for 2-forms which are gauge invariant. A natural quantity
to consider is the norm of $\ymF$, defined by the Killing-Cartan form
$K_{ij}$ on the Lie-algebra of $G$ and the (contravariant) $\twoeps$
on the horizon:
\begin{equation}\label{modf}
\modF\,\, := \left[ (\twoeps\into \ymF)^i (\twoeps\into \ymF)^j\,
K_{ij} \right]^{1\over 2}
\end{equation}
(Although the contravariant $\twoeps$ is ambiguous up to terms of the
type $\ell^{[a}V^{b]}$ where $V^a$ is any vector field tangential to
$\Delta$, this ambiguity does not affect $\modF$ because of
(\ref{pbackymf}). The norm of $\dual \ymF$ is defined
analogously. These two quantities are gauge invariant and allow us to
define the electric and magnetic Yang-Mills charges of the horizon:
\begin{equation}\label{ymcharge}
    Q^{YM}_{\Delta} :\hateq -\frac{1}{4\pi}
    \oint_{S_{\Delta}}\modstarF\, \twoeps
    \qquad \qquad \mbox{and} \qquad \qquad   P^{YM}_{\Delta} :\hateq -\frac{1}{4\pi}
    \oint_{S_{\Delta}}  \modF \, \twoeps \, .
\end{equation}
Recall that the unusual signs in the definitions of the charges arise
due to the orientation of the $S_{\Delta}$ --- the normal to the two
sphere is inward pointing.  In Maxwell theory, the magnetic charge is
zero unless we consider either connections on non-trivial bundles or
allow `wire singularities'. As is well known, this is not true for
Yang-Mills theory: the magnetic charge can be non-zero even if we
restrict attention to smooth fields on a trivial bundle.

We would now like to verify that the charges defined in
(\ref{ymcharge}) are independent of the cross section of the horizon
$S_{\Delta}$ on which the integration is performed. The isolated
horizon boundary conditions guarantee this is the case. First, recall
the geometric identity
\begin{equation}\label{lieymf}
\Lie_{\ell}\pback{\ymF} \hateq
    \ell \into \pback{\ymD\ymF} - [(\ell \cdot \ymA), \pback{\ymF}]  +
\pback{\ymD(\ell \into \ymF)}.
\end{equation}
A similar expression for $\dual \ymF$ is also true. The first term on
the right hand side vanishes due to the field equations and the third
term is zero due to the previous restriction on $\ymF$,
(\ref{pbackymf}). Therefore at the isolated horizon,
\begin{equation}\label{lieymfhor}
    \Lie_{\ell}\pback{\ymF} \hateq - [(\ell \cdot \ymA),\ymF]
    \qquad \mbox{and} \qquad
    \Lie_{\ell}\pback{\dual \ymF} \hateq - [(\ell \cdot \ymA),
    \dual\ymF]\, .
\end{equation}
In the Maxwell case, $\emF$ and $\dual \emF$ are Lie dragged by $\ell$.
However, for non-Abelian fields, this is not a gauge invariant
statement; the terms on the right hand sides of (\ref{lieymfhor}) are
necessary for gauge invariance.  Although the field strength and its
dual are not Lie dragged along $\ell^a$, recalling that $\Lie_{\ell}
\twoeps \hateq 0$ and using the cyclic property of the trace, it is
straightforward to demonstrate that their norms are:
\begin{equation} \label{liemodf}
    \Lie_{\ell}\modF\,\hateq 0
    \qquad \mbox{and} \qquad
    \Lie_{\ell}\modstarF\, \hateq 0.
\end{equation}
This result, along with (\ref{pbackymf}), guarantees that the charges
$Q^{YM}_{\Delta}$ and $P^{YM}_{\Delta}$ are independent of the choice
of cross section $S_{\Delta}$ of the horizon.

Let us consider the remaining components of the Yang-Mills field. The
boundary conditions place no restrictions on $\ymF_{ab}\, n^a m^b$ and
$\dual \ymF_{ab}\, n^a m^b$ at all.  As in the electromagnetic case,
these components describe the radiation flowing along the horizon. The
isolated horizon boundary conditions allow radiation arbitrarily close
to --- and even at --- the horizon, provided none crosses it.

We have so far restricted our attention to the field strength and its
dual.  However, in the action principle and phase space, the basic
variable will be the Yang-Mills connection $\ymA$.  Let us begin with
the definition of the Yang-Mills equivalent of the electric potential.
Recall that, given an $\ell$, the electric potential was defined in
Section \ref{s2.D} as $\Phi_{(\ell)} \hateq -(\ell \cdot \emA)$. This
definition is not appropriate in the Yang-Mills case since the
resulting potential has a free internal index and is therefore not
gauge invariant. Instead, we define the Yang-Mills potential,
$\ymphi_{(\ell)}$ to be negative the norm of $(\ell \cdot \ymA)$:
\begin{equation}\label{ymphidef}
\ymphi_{(\ell)} :\hateq - \mid\! (\ell \cdot \ymA) \!\mid.
\end{equation}
This gives us a gauge invariant potential at the horizon.

As in Maxwell theory, we need to constrain the form of $\ymA$ at the
horizon.  Several considerations motivate our choice of these boundary
conditions.  First they must be chosen so that the action principle is
well defined. Second, if the gauge group is U(1) the boundary
conditions should reduce to those given in Section \ref{s2.D} for the
electromagnetic field. Finally, we should be able to show that the
Yang-Mills electric potential is constant on the horizon.  These
considerations suggest the following definition:

\medskip
\noindent\textbf{Definition 6}: The connection $\ymA$ will be said to
be in a \textit{gauge adapted to the isolated horizon} $(\Delta, [\l])$
if it satisfies the following two conditions
\begin{itemize}
\item (i) The Yang-Mills potential is constant on the horizon
\begin{equation}\label{ymphiconst}
    d\ymphi_{(\ell)} \hateq 0
\end{equation}
\item (ii)
The dual of the field strength $(\dual \ymF)$ and $(\ell \cdot \ymA)$
point in the same Lie algebra direction,
\begin{equation}\label{samedir}
    (\ell \cdot \ymA)^i \propto (\twoeps \into \dual\ymF)^i
\end{equation}
on the horizon.
\end{itemize}

These boundary conditions satisfy the requirements discussed above.
First, it is straightforward to show that in the U(1) case, condition
(i) is equivalent to requiring $\Lie_{\ell} \emA \hateq 0$ and (ii) is
redundant. Second, as we shall see in Section \ref{s6.B}, these
boundary conditions are also sufficient to make the variational
principle well defined.

It is not difficult to show that these conditions can always be
satisfied. The remaining gauge freedom is simply $\ymA \rightarrow
g^{-1} \ymA g + g^{-1}\partial g$, where $g$ satisfies $\Lie_{\ell} g
\hateq 0$.

\subsection{Action and Phase Space}
\label{s6.B}

 In this section we will consider the first order action
for Einstein-Yang-Mills theory on the manifold $\man$ described in
Section \ref{s3.B}.  The basic fields will consist of the triplet
$(e_a^I,A_{aI}{}^J,\ymA_a^i)$, where $e_a^I$ and ${A_{aI}}^J$ are the
tetrad and Lorentz connection and $\ymA_a^i$ is the Yang-Mills
connection.  The gravitational fields, $e_a^I$ and ${A_{aI}}^J$ satisfy
the same boundary conditions (at $\Delta$ and infinity) as in Section
\ref{s3.B}.  Furthermore, we require the Yang-Mills fields to be in a
gauge adapted to the horizon and assume they fall off
sufficiently fast at infinity.%
\footnote{More specifically we require the fall off conditions on the
Yang-Mills connection to be such that all integrals, in particular the
symplectic structure, are finite and yet the asymptotic electric and
magnetic charges are not forced to vanish. While it is not trivial to
meet these conditions (for examples the conditions used in \cite{sw,cs}
appear not to lead to a well-defined symplectic structure) they can be
met. However, as in the rest of the paper, for brevity, we will not
spell out the boundary conditions at infinity in detail.}
The Einstein-Yang-Mills action is
\begin{equation}\label{ymaction}
S (e, A, \emA) = \frac{-1}{16\pi G} \int_{\man} \Sigma^{IJ}\wedge
F_{IJ} +\frac{1}{16\pi G} \int_{\tau_\infty} \Sigma^{IJ} \wedge A_{IJ}
    -\frac{1}{8\pi}\int_{\man} \Tr[\ymF\wedge \dual\ymF]\, .
\end{equation}
The gravitational part of the action has previously been discussed in
Section \ref{s3}, therefore we shall only consider in detail the
Yang-Mills terms and verify the variational principle is well defined.
Taking into account the results of Section \ref{s3}, a variation of the
action can be expressed as
\begin{equation} \label{ymvar}
    \delta S(e,A,\emA)  =
    \int_{\man} \mbox{Equations of Motion} \cdot\delta \phi
    -\frac{1}{4\pi} \int_{\Delta} \Tr[\delta\ymA \wedge \dual\ymF].
\end{equation}
We must demonstrate the boundary term at $\Delta$ vanishes due to the
conditions imposed on the Yang-Mills fields.  Using (\ref{pbackymf})
and (\ref{samedir}), one can show that the trace in (\ref{ymvar}) can
be replaced by a product of norms:
\[
    \Tr[\delta\ymA \wedge \dual \ymF] =
    (\delta\ymphi_{(\l)} - c_{\delta}\ymphi_{(\l)})
    \modstarF\,  {}^{\Delta}\epsilon \, ,
\]
where ${}^{\Delta}\epsilon = n \wedge \twoeps$ is the volume form on
$\Delta$ and $\delta \ell = c_{\delta}\ell$.

In the action principle, variations are performed keeping data fixed at
the initial and final slices. In particular $\delta \ymphi_{(\ell)}$
and $c_{\delta}$ vanish there. However, the boundary conditions
guarantee that $\ymphi_{(\ell)}$, and hence its variation, is constant
on $\Delta$. Since, $\delta \ymphi_{(\ell)}$ and $c_{\delta}$ vanish on
the initial cross section of the horizon and are constant, it follows
that $\delta \ymphi_{(\ell)} \hateq 0$ and $c_{\delta} \hateq 0$.
Therefore, the Yang-Mills horizon boundary term vanishes; the action
principle is well defined in the presence of Yang-Mills fields.  As in
the Einstein-Maxwell case, boundary conditions played a crucial role in
demonstrating the viability of the action.

We now wish to construct the covariant phase space and symplectic
structure.  As before, points in covariant phase space $\Gamma$ will
consist of histories which satisfy the Einstein-Yang-Mills field
equations, appropriate falloff conditions at infinity and the isolated
horizon boundary conditions at $\Delta$.  Before proceeding further, we
shall once again need to introduce an additional field at the horizon.
This can be regarded as `potential' for the Yang-Mills potential
$\ymphi$. Given any point $(e,A,\emA)$ in the phase space $\Gamma$, let
us define the scalar field $\chi$ on $\Delta$ as follows: i) ${\cal
L}_{\l} \chi =- \ymphi_{\ls}$; and ii)  $\chi$ vanishes on $S^-$, the
intersection of $M^-$ with $\Delta$.  These conditions are identical to
those imposed in the Maxwell case.

Once again, we take second variations of the action in order to obtain
a symplectic structure.  Since the gravitational terms are exactly the
same in Section \ref{s3}, we shall only describe in detail the
Yang-Mills part of the symplectic structure. The second variation of
the action (\ref{ymaction}) yields the following symplectic current:
\begin{equation}\label{ymsymcur}
J(\gamma; \delta_1,\delta_2)= J_{\rm grav} -
    \frac{1}{4\pi} \Tr[\delta_1 \dual \ymF \wedge \delta_2 \ymA
        -\delta_2 \dual \ymF \wedge \delta_1 \ymA ].
\end{equation}
Using the fact that the field equations and linearized field equations
are satisfied, one can directly verify that $J(\gamma; \delta_1,
\delta_2)$ is indeed a closed 3-form.  We again choose the spacetime
region of interest, $\tilde{\man}$, to be that part of the spacetime
$\man$ bounded by $M_1$, $M_2$, infinity and a portion $\tilde\Delta$
of the isolated horizon. Integrating $dJ$ over $\tilde{\man}$ and using
asymptotic falloff conditions, we obtain
\[ (\int_{M_1} + \int_{M_2} + \int_{\tilde\Delta}) J(\gamma; \delta
q_1,\delta_2) = 0. \]
The integral of $J$ over $\tilde{\Delta}$ does not vanish but, as in
Section \ref{s3}, the pull back of $J$ to $\Delta$ is exact. Therefore,
we can express the integral over $\tilde{\Delta}$ of $J$ as integrals
over the initial and final 2-spheres $S_1$ and $S_2$.  Using these
results, and keeping track of orientations, we obtain the symplectic
structure:
\begin{eqnarray}\label{ymsym}
   \lefteqn{\Omega|_{\gamma} \, (\delta_1 ,\delta_2 ) =} \nonumber\\
   && -\frac{1}{16\pi G}\int_{M} [\delta_1 \Sigma^{IJ}\wedge \delta_2 A_{IJ}
        -\delta_2 \Sigma^{IJ} \wedge \delta_1 A_{IJ}] \,
    + \frac{1}{8 \pi G}\oint_S [\delta_1\,\, \twoeps \delta_2\psi
        - \delta_2\,\, \twoeps \delta_1\psi] \nonumber\\ &&
    -\frac{1}{4\pi } \int_{M} [\delta_1 \dual \ymF \wedge \delta_2 \ymA
        -\delta_2 \dual \ymF \wedge \delta_1 \ymA] \,
    +\frac{1}{4\pi} \oint_S [\delta_1\,\, (\modstarF \twoeps) \delta_2\chi
    - \delta_2\,\,(\modstarF \twoeps) \delta_1 \chi] \, .
\end{eqnarray}
Full use of the isolated horizon boundary conditions has been made in
obtaining this symplectic structure. In particular, to obtain the given
form of the Yang-Mills surface term, we have used the fact that $(\ell
\cdot \ymA)$ and $(\twoeps \into \dual\ymF)$ point in the same
direction in the Lie algebra.  Using field equations one can directly
verify that the right side of (\ref{ymsym}) is independent of the
choice of the partial Cauchy surface; the symplectic structure is
`conserved'. We will use $(\Gamma, \Omega)$ as our covariant phase
space.

\subsection{Hamiltonian and First Law}

In this subsection we will generalize the arguments of Section \ref{s5}
to obtain an expression for the energy of the horizon in
Einstein-Yang-Mills theory.  To do so, we must specify a time evolution
vector field $t^a$.  As before, we require $t^a$ to be a member of the
preferred equivalence class $[\ell^a]$ at the horizon (this again
requires restriction to non-rotating isolated horizons) and approach
unit time translation asymptotically. Given $t^a$ we can calculate the
infinitesimal generator of time evolution, $\delta_t = (\Lie_t e,
\Lie_t A, \Lie_t \ymA)$, and determine whether it is Hamiltonian.
Recall that $\delta_t$ is Hamiltonian if and only if the 1-form $X_t$
on the phase space defined by
\begin{equation}\label{ymdefx}
    X_t(\delta):= \Omega(\delta, \delta_{t})
\end{equation}
is closed.  Let us calculate $X_t$.  The gravitational part will be
identical to the expression obtained in Section \ref{s5}, therefore we
shall concentrate on the Yang-Mills terms.  As with the Maxwell field,
the Lie derivatives of the Yang-Mills fields can be re-expressed using
the following identities:
\begin{equation}\label{ymliet}
\Lie_t \ymA =  t\into \ymF + \ymD(t\cdot \ymA)  \qquad \qquad
\Lie_t\dual\ymF = t\into(\ymD\dual\ymF) - [\ell \cdot \ymA,\dual \ymF]
+\ymD(t\into \dual \ymF)\, .
\end{equation}
Making use of these expressions, the field equations satisfied by
$(e,A, \ymA)$ and the linearized field equations for $\delta$, we
obtain the required expression for $X_t$:
\begin{equation}\label{ymxt}
    X_t(\delta) =
    \frac{-1}{16\pi G} \int_{\partial M} (t\cdot A) \delta\Sigma
        - (t\into \Sigma) \wedge\delta A
    - \frac{1}{4\pi}\int_{\partial M}\Tr[(t\cdot \ymA) \delta(\dual \ymF)
        - (t\into \dual \ymF)\wedge\delta\ymA].
\end{equation}
As before the expression involves integrals \textit{only} over the
2-sphere boundaries $S_\infty$ and $S_\Delta$ of $M$; there is no
volume term.  The gravitational terms yield $\delta E_{\rm ADM}^{t}$ at
infinity and $-(1/8\pi G)\kappa_{(t)} \delta a_{\Delta}$ at the
horizon.  The Yang Mills term at infinity vanishes due to fall-off
conditions, therefore we need only calculate the Yang-Mills
contribution at the horizon.  This is composed of two terms, the second
of which vanishes due to the restriction (\ref{pbackymf}) which
guarantees $\ell \into \pback{\dual \ymF} \hateq 0$. Since we are in a
gauge adapted to the horizon, $(\ell \cdot \ymA)$ and $\dual \ymF$
point in the same internal direction.  This allows us to replace the
trace in the first term by norms:
\[ \Tr[(t\cdot \ymA) \delta(\dual \ymF)] = -\ymphi_{(t)} \delta(\modstarF
\twoeps) \]
and guarantees that $\ymphi_{(t)}$ is constant on the horizon.  Making
use of the definition of Yang-Mills electric charge, (\ref{ymcharge}),
we obtain
\[
    X_t(\delta) =
    -\frac{1}{8\pi G}\kappa_{(t)} \delta a_{\Delta}
    -\ymphi_{(t)} \delta Q^{\YM}_{\Delta}
    +\delta E_{\rm ADM}^t.
\]
Recall that the necessary and sufficient condition for the existence of
a Hamiltonian is that $X_t$ be closed. Clearly, this is equivalent to
\begin{equation}\label{ymiff}
\frac{1}{8\pi G}\,\dd \kappa_{(t)} \wedge \dd a_{\Delta} +
\dd\ymphi_{(t)} \wedge \dd Q^{YM}_{\Delta} = 0,
\end{equation}
Once again, we conclude that the surface gravity $\kappa_{(t)}$ and the
Yang-Mills potential $\ymphi_{(t)}$ at the horizon defined by $t^a$ can
depend \textit{only} upon the area $a_{\Delta}$ and charge
$Q^{\YM}_{\Delta}$ of the horizon. Finally, (\ref{ymiff}) is also the
necessary and sufficient condition that there exist a function
$E_{\Delta}^{t}$, also only of $a_{\Delta}$ and $Q^{\YM}_{\Delta}$ such
that
\begin{equation}\label{ym1law}
    \delta E_{\Delta}^{t} = \frac{1}{8\pi G}\,\kappa_{(t)}
    \delta a_{\Delta} + \ymphi_{(t)} \delta Q^{\YM}_{\Delta}.
\end{equation}
As before, $E_{\Delta}^{t}$ is interpreted as the horizon energy
defined by the time translation $t^{a}$.  We conclude that the vector
field $\delta_t$ is Hamiltonian if and only if the first law,
(\ref{ym1law}), holds.

We will conclude this section with a few remarks.

i) The derivation of the first law and its final form are completely
analogous to those in the Einstein-Maxwell theory.  By contrast, in the
discussion of the first law for undistorted isolated horizons of
\cite{cs}, certain restrictions were imposed on the permissible
variations $\delta$ in the Einstein-Yang-Mills case. In our treatment,
subtleties arise only in the definition of a canonical horizon mass
(see Section \ref{s7.C}) rather than the discussion of the first law
itself.

ii) Although the Yang-Mills magnetic charge $P_\Delta^{\rm YM}$ will
generically not be zero, no term involving $\delta P^{\rm YM}_\Delta$
arises in the first law.

iii) How does our result compare with those previously available
\cite{sw,hs}?  In \cite{hs} the first law for Yang-Mills fields is
proved for globally stationary spacetimes and small perturbations from
one such space-time to another.  Assuming the Yang-Mills fields fall
off sufficiently fast at infinity, (in the non-rotating case) the first
law of \cite{hs} then reads
\[
    \delta M = \frac{1}{8 \pi G} \kappa \delta a
    + \int_{\mathrm{Hor}} \Tr[\phi \, \delta \dual F]\, .
\]
Here, $M$ is the ADM mass evaluated at infinity, while all terms on
the right hand side are evaluated at the horizon. Due to a different
gauge choice at the horizon, the authors define a Lie algebra valued
potential $\phi$ and leave the `$\Phi \delta Q$' term inside an
integral. However, the general form of this first law is the same as
ours.  In this sense, our framework generalizes the results of
\cite{hs} to non-static contexts.

In \cite{sw}, the first law is proved for globally stationary
spacetimes and arbitrary small departures therefrom. However, there
are a number of important differences between these results and the
ones obtained in this paper.  In the non-rotating case, the first law
of \cite{sw} reads
\[
    \delta M_{\mathrm{ADM}} + V \delta Q =
    \frac{1}{8 \pi G} \kappa \delta a
\]
where $V$ and $Q$ are the Yang-Mills potential and charge evaluated at
\textit{infinity} while $\kappa$ and $a$ are of course evaluated at the
horizon.  Because of the non-Abelian nature of the Yang-Mills field, unlike
in the Maxwell case, the charge $Q$ evaluated at infinity is now quite
different from the charge evaluated at the horizon and, as in the
Maxwell theory, the potential $V$ evaluated at infinity has no direct
bearing on the potential at the horizon.  Furthermore, that calculation
makes an essential use of the bifurcation 2-sphere and all fields are
required to be smooth there.  This restriction implies that the
Yang-Mills potential at the horizon \textit{vanishes}.  (The same is
true if one restricts the analysis of \cite{sw} to the Maxwell case.)

The first law derived in the isolated horizon framework is valid also
in presence of radiation in the exterior space-time region and makes no
reference to the bifurcation 2-sphere.  (Although we restricted
ourselves to the non-rotating case, rotation has been incorporated in
this framework in \cite{abl2}.)  Furthermore, it has the aesthetically
pleasing feature that all quantities that appear in (\ref{ym1law}) ---
including the energy $E^t_\Delta$, the potential $\ymphi_{(t)}$ and the
charge $Q^{\YM}_\Delta$ --- are evaluated at the horizon.  In
particular, one can now meaningfully consider the physical process
version in which one does an experiment \textit{at} the horizon by
dropping a test particle/field and changing the \textit{horizon charge}
infinitesimally.  More generally (\ref{ym1law}) is genuinely a law
governing the mechanics of the \textit{horizon}.

\section{Horizon Mass}
\label{s7}

For notational simplicity, we will say that a (live) vector field $t^a$
is \textit{permissible} if it gives rise to a Hamiltonian evolution. We
saw in Sections \ref{s5} and \ref{s6} that each permissible vector
field $t^a$ defines a horizon energy $E^t_{\Delta}$. In the phase space
framework, $E^t_{\Delta}$ has a direct interpretation: it is the
surface term at the horizon in the expression of the Hamiltonian
generating the $t^a$-evolution. However, in many physical applications
--- such as the study of black hole mergers --- one is interested in
properties of a \textit{specific} space-time, rather than the full
phase space. Then, it is useful to have at one's disposal a canonical
notion of energy, the analog of the ADM energy in the rest frame at
infinity. This quantity could then be interpreted as the horizon mass.
In this section, we will introduce this notion in detail. The
discussion is divided into three parts. In the first, we consider the
Einstein-Maxwell theory; in the second, we discuss dilatonic couplings
\cite{ac}; and, in the third, we analyze the Einstein-Yang-Mills
system.

\subsection{Einstein-Maxwell theory}
\label{s7.A}

In Section \ref{s5} we showed that $t^a$ is permissible if and only if
(\ref{iff}) holds on the phase space.  We will now construct a large
family of permissible evolution fields $t^a$. Fix any regular function
$\kappa_o$ of two variables $a_\Delta$ and $Q_\Delta$. Then, given any
point $\gamma \equiv (e,A,\emA)$ of $\Gamma$, we define (the boundary
value of) the vector field $t^a$ as follows. Consider the vector field
$\l^a$ on $\Delta$ defined by the tetrad, $\l^a = e^a_I \l^I$, and
denote by $\kappa_{\ls}$ the surface gravity associated with it. Then,
$\kappa_o = c \kappa_{\ls}$ for some constant $c$. Let us set $t^a = c
\l^a$. Repeating this procedure at each phase-space point $\gamma$, we
obtain a live vector field $t^a$ with $\kappa_{(t)} = \kappa_o$. (The
resulting $c$ will be constant on $\Delta$ but a function on the phase
space.)  Next, consider the electro-magnetic potential, which is
guaranteed to be constant on $\Delta$ by our boundary conditions but
whose value at any phase space point is so far completely free. We will
now use (\ref{iff}) to fix it. Equation (\ref{iff}) implies:
\[ \frac{\partial \kappa_{(t)}}{\partial Q_{\Delta}} =
\frac{\partial \Phi_{(t)}}{\partial a_{\Delta}}. \]
Since $\kappa_{(t)} = \kappa_o$ is known, we can simply integrate the
equation for $\Phi_{(t)}$ as a function of $a_\Delta$ and $Q_\Delta$.
Furthermore, the solution is unique if we impose the physical condition
that $\Phi_{(t)}$ should vanish whenever $Q_\Delta =0$. Thus, starting
from any regular function $\kappa_o$ of $a_\Delta$ and $Q_\Delta$, we
have obtained a permissible evolution field $t^a$. Conversely, it is
easy to verify that every permissible vector field arises via this
construction. There is clearly a very large family of such live vector
fields.

An obvious question is if there is a `canonical' or `natural' choice of
$t^a$? We will now show that the answer is in the affirmative. Recall
that, in the Einstein-Maxwell theory, there is precisely a 2-parameter
family of \textit{globally static} solutions admitting horizons: the
Reissner-Nordstr\"om family. (Since $\emA$ is required to be a globally
defined connection on a trivial $U(1)$ bundle, the magnetic charge is
zero on the entire phase space.) Let us focus on this family. Denote by
$\xi^a$ the static Killing field which is \textit{unit} at infinity.
Its surface gravity is a specific function of $a_\Delta$ and
$Q_\Delta$:
\[\kappa_{(\xi)} = \frac{1}{2R_\Delta}
\left( 1- \frac{GQ_\Delta^2}{R^2_\Delta}\right) \, . \]
As before, $R_\Delta$ is the horizon radius, defined by $a_\Delta =
4\pi R^2_\Delta$.  We can therefore use $\kappa_{(\xi)}$ in place of
$\kappa_o$ in the above construction.  The resulting permissible, live
vector field $t^a_o$ agrees with $\xi^a$ on the horizon of every static
solution. This property is satisfied \textit{only if} we set $\kappa_o
= \kappa_{(\xi)}$.

Next, we can `integrate' (\ref{1law}) to obtain the horizon energy
$E_\Delta^{t_o}$. Although a priori there is the freedom to add a
constant, we can fix it by requiring that the energy vanish as
$a_\Delta$ and $Q_\Delta$ tend to zero. Indeed, we have no choice in
this since one cannot construct a quantity with dimensions of mass from
the fundamental constants that appear in the Einstein-Maxwell theory.
(Einstein-Yang-Mills theory does admit such a constant and we will
see in Section \ref{s7.C} that it leads to an interesting modification
of the situation discussed here.) Let us define the horizon mass via
\[ M_\Delta = E^{t_o}_\Delta. \]
To justify this definition, let us begin by restricting to static
solutions. In each static solution, we are free to extend $t^a_o$ such
that it coincides with the Killing field $\xi^a$. General symplectic
arguments imply that, on any connected component of the space of static
solutions, the numerical value of the total Hamiltonian, generating
evolution along $\xi^a$, must be constant (see, e.g., \cite{abf2}.) In
the Einstein-Maxwell case, there is a single connected component and,
by the dimensional argument given above, the numerical value of the
Hamiltonian must vanish on it. Hence, from (\ref{ham}) it follows that,
on any static solution,
\[ H_{t_o} = M_{\rm ADM} - M_\Delta = 0. \]
On a general solution, of course, $M_{\rm ADM}$ would be greater than
$M_\Delta$, the difference being equal to the energy in radiation. If
the horizon is complete in the future and time-like infinity $i^+$
satisfies certain regularity conditions, as in \cite{abf2} one can
argue that the difference is precisely the total energy radiated across
${\cal I}^+$ and hence $M_\Delta$ equals the future limit of the Bondi
mass. These considerations support our interpretation of $M_\Delta$ as
the horizon mass.

Finally, since we now have a canonical evolution field $t^a_o$, we can
drop the suffix $t$ on surface gravity and electromagnetic potential
and write the first law (\ref{1law}) in the more familiar form:
\[ \delta M_\Delta = \frac{1}{8\pi G}\, \kappa \delta a_\Delta
+ \Phi \delta Q_\Delta \, . \]
In contrast to treatments based on static space-times, the quantities
that enter this law are all defined \textit{at} the horizon. Therefore,
as pointed out in \cite{abf2}, it is now possible to interpret this law
also in the `active' sense where one considers physical processes which
increase the area and the charge of a given horizon. To our knowledge,
the standard proofs of this physical version \cite{mh,w2} are not
applicable to processes in which the background has non-zero electric
charge and the process changes it infinitesimally.

We will conclude with a few remarks.

i) In the above discussion, the permissible evolution field $t_o^a$ was
constructed by setting $t^a_o = c \l^a$ where $c$ is given by $\kappa_o
\equiv ({1}/{2R_\Delta}) (1 - G(Q_\Delta/R_\Delta)^2) = c
\kappa_{\ls}$. For $c$ to be well-defined, it is necessary that
$\kappa_o$ vanishes whenever $\kappa_{\ls}$ does. Therefore, for the
mass to be well-defined, we must excise those points from the phase
space at which $\kappa_{\ls}$ vanishes but $\kappa_o$ does not.
However, this is not a serious limitation. In particular, we still
retain all static solutions \textit{including the extremal ones} at
which $\kappa_o$ vanishes.

ii) Since we have a specific $\kappa_o$, we can use (\ref{iff}) to
obtain the corresponding electrostatic potential: $\Phi_o =
Q_\Delta/R_\Delta$. Furthermore, by integrating (\ref{1law}) it is easy
to express $M_\Delta$ explicitly in terms of the horizon parameters:
\begin{equation} \label{mass}
    M_\Delta =  \frac{1}{4\pi G}\,\kappa a_\Delta + \Phi Q_\Delta
    =\frac{R_\Delta}{2G}
    \left(1 + \frac{GQ_\Delta^2}{R_\Delta^2}\right)\, .
\end{equation}
Thus, the functional dependence of $M_\Delta$ on the horizon
parameters at any point of the phase space is the same as in static
space-times.  Note that this is a \textit{result} of the framework,
not an assumption. Its derivation involved two distinct steps. First,
and most importantly, the first law (\ref{1law}) arose as a necessary
and sufficient condition for the existence of a consistent Hamiltonian
framework. Second, the freedom in $t^a$ was exploited in order to
construct the preferred, permissible evolution field $t_o^a$. It is
quite significant that $M_\Delta$ can be expressed so simply using
just the parameters defined \textit{locally} at the horizon even when
there is radiation arbitrarily close to it.  This fact is likely to
play an important role in the problem of extracting physics in the
strong field regimes from numerical simulations of black hole
collisions \cite{abl1}. It is important to notice that although we
made use of our knowledge of static solutions to arrive at a canonical
$t^a_o$ and the mass function $M_\Delta$, the final result
(\ref{mass}) \textit{makes no reference} to these
solutions. $M_\Delta$ is a simple function of the parameters which can
be directly computed from the geometry of any one isolated horizon.

iii) In the earlier work \cite{abf2} on undistorted horizons, one
restricted oneself to the preferred evolution field $t^a_o$ from the
very beginning (although this vector field was selected using a
different but equivalent procedure).%
\footnote{This strategy seems to have generated a misunderstanding
(see, e.g.,\cite{w3}) that the first law was obtained in
\cite{abf2,ac,cs} merely by identifying the parameters labeling a
general isolated horizon with those of static horizons and then using
the Smarr formulas available in the static context. This was not the
case. Rather, static solutions were used \textit{only} to select the
appropriate normalization of the evolution vector field $t^a$ at the
horizon. The Hamiltonian framework was then used to define the horizon
mass without any reference to Smarr formulas. As in this section, the
mass was then \textit{shown} to reproduce the Smarr-type formulas on
general horizons.}
The a priori freedom in the choice of a permissible $t^a$ was not
discussed and the first law appeared only in the more familiar form,
given above.

\subsection{Dilatonic coupling}
\label{s7.B}

The Einstein-Maxwell-dilaton system was studied in some detail in the
undistorted case in \cite{ac}. We will revisit it here in the more
general context considered in this paper because it brings out a
subtlety in the definition of the horizon mass $M_\Delta$ and the
associated first law.

The dilaton is a scalar field $\phi$ which can couple to the Maxwell
field in a non-standard fashion.  The coupling is governed by a
constant $\alpha$.  If $\alpha$=$0$, one obtains the standard
Einstein-Maxwell-Klein-Gordon theory and the situation then is
completely analogous to the Einstein-Maxwell theory considered above.
If $\alpha$=$1$, the theory represents the low energy limit of string
theory.  In this case, there are some interesting differences from the
Einstein-Maxwell theory considered in this paper.  To bring out these
differences, in this subsection we will set $\alpha$=$1$.  (The
situation for a general value of $\alpha$ is discussed in \cite{ac}
where one can also find details on the material summarized below.)

In the standard formulation, the theory has three charges, all defined
at infinity; the ADM mass $M_{\rm ADM}$, the usual electric charge
$Q_\infty$ and another charge $\tilde{Q}_\infty$:
\[Q_\infty = \frac{1}{4\pi}\, \oint_{S_\infty}\, \dual\emF\,
\quad {\rm and} \quad \tilde{Q}_\infty = \frac{1}{4\pi}\,
\oint_{S_\infty}\, e^{-2\phi}\,\, \dual\emF \, .
\]
$\tilde{Q}$ is conserved in space-time (i.e. its value does not change
if the 2-sphere of integration is deformed) while $Q$ is not. From the
perspective of the isolated horizons, it is more useful to use
$a_\Delta, Q_\Delta, \tilde{Q}_\Delta$ as the basic charges%
\footnote{In the undistorted case, the dilaton is constant on $\Delta$
and hence we can replace $Q_\Delta$ by $\phi_\Delta$ as in \cite{ac}.}:
\[Q_\Delta = \frac{1}{4\pi}\, \oint_{S_\Delta}\, \dual\emF\, , \quad
{\rm and} \quad \tilde{Q}_\Delta = \frac{1}{4\pi}\, \oint_{S_\Delta}\,
e^{-2\phi} \dual\emF \, .
\]
Although the standard electric charge is not conserved in space-time,
it \textit{is} conserved along $\Delta$ whence $Q_\Delta$ is
well-defined.

It is straightforward to extend the construction of the phase space to
include the dilaton. The only difference is that the charge $Q$ in
equations (\ref{xtbc}) - (\ref{1law}) is replaced by $\tilde{Q}$.  With
this minor change, the discussion of the first part of Section
\ref{s7.A} is also unaffected. Thus, given any function $\kappa_o$ of
$a_\Delta$ and $\tilde{Q}_\Delta$, we can construct a permissible,
(live) evolution field $t^a$.

The difference arises in the next step where we constructed a preferred
$t_o^a$. With the dilatonic coupling, the theory has a unique
\cite{masood} \textit{three} parameter family of static solutions which
can be labeled by $(a_\Delta, Q_\Delta, \tilde{Q}_\Delta )$. As in the
Reissner Nordstr\"om family, these solutions are spherically symmetric.
In terms of these parameters, the surface gravity $\kappa_{(\xi)}$ of
the static Killing field which is unit at infinity is given by:
\[\kappa_{(\xi)} = \frac{1}{2R_\Delta} \left[ 1 + 2G \frac{Q_\Delta
\tilde Q_\Delta}{R_\Delta^2}\right] \left[ 1 - 2G \frac{Q_\Delta \tilde
Q_\Delta}{R_\Delta^2}\right]^{-\frac{1}{2}} \,\, .\]
The problem in the construction of the preferred $t_o^a$ is that we
need a function $\kappa_o$ which depends only on $a_\Delta$ and
$\tilde{Q}_\Delta$.  Therefore, we can no longer set $\kappa_o =
\kappa_{(\xi)}$ on the entire phase space because $\kappa_{(\xi)}$
depends on all three horizon parameters.

To extract the mass function $M_\Delta$ on the phase space, we can
proceed as follows.  Let us foliate $\Gamma$ by $Q_\Delta = {\rm
const}$ surfaces.  On each leaf, $\kappa_{(\xi)}$ trivially depends
only on $a_\Delta$ and $\tilde{Q}_\Delta$ and so we can set $\kappa_o
= \kappa_{(\xi)}$.  Therefore, by the procedure outlined in Section
\ref{s7.A}, we obtain a (live) vector field $t_o^a$ and can define the
mass $M_\Delta (\gamma) = E_{\Delta}^{t_o}(\gamma)$ for all points
$\gamma$ on this leaf.  Repeating this procedure for each leaf, we
obtain a live vector field $t_o^a$ and a mass function $M_\Delta$
everywhere on $\Gamma$.  However, the surface gravity $\kappa_{(t_o)}$
now depends on all three parameters, rather than just $a_\Delta$ and
$\tilde{Q}_\Delta$.  Therefore, the first law (\ref{1law}) cannot hold
for arbitrary variations $\delta$ and consequently $\delta_{t_o}$
fails to be a Hamiltonian vector field.  Put differently, although
there is a multitude of permissible, live vector fields, each leading
to a first law, none of them can coincide with the Killing field
$\xi^a$ (which is unit at infinity) on \textit{all} static solutions.
This is a significant departure from the Einstein-Maxwell case
considered above.

Nonetheless, (modulo the caveat discussed in the first remark at the
end of Section \ref{s7.A}) the above procedure does provide us with a
well-defined mass function $M_\Delta$ on the entire phase space which
can be expressed in terms of the horizon parameters as
\[M_\Delta = \frac{R_\Delta}{2G} \left[ 1 - 2G \frac{Q_\Delta
\tilde{Q}_\Delta}{R_\Delta^2}\right]^{-\frac{1}{2}} \, .\]
It equals $M_{\rm ADM}$ in static space-times and has other properties
which motivated our interpretation of $M_\Delta$ as the horizon mass in
the Einstein-Maxwell case. Since this function is well-defined on the
entire phase space, we can simply vary it and express the result in
terms of the horizon parameters. The result is:
\[\delta{M}_\Delta = \frac{1}{8\pi G}\, \kappa \delta{a_\Delta}
+\hat\Phi \delta \hat{Q}_\Delta\]
where $\kappa = \kappa_{(t_o)}$, $\hat\Phi^2 = (Q_\Delta
\tilde{Q}_\Delta/R_\Delta^2)$ and $\hat{Q}_{\Delta}^2 = Q_\Delta
\tilde{Q}_\Delta$. Thus, although there is still a first law in terms
of $t^a_o$ and $M_\Delta$, it does not have the canonical form
(\ref{1law}) because $t_o^a$ is not a permissible vector field.  More
generally, in theories with multiple scalar fields \cite{gk}, if one
focuses only on static sectors, one obtains similar `non-standard'
forms of the first law with work terms involving scalar fields. This
reflects the fact that there is no permissible vector field $t^{a}$,
defined for all points of the phase space, which coincides with the
properly normalized Killing field on \textit{all} static solutions.  In
the undistorted case, the analysis
 was carried out only in terms of the vector field $t^a_o$ and
the horizon mass $M_\Delta$ \cite{ac}. The resulting first law had the
above form.

Alternatively, one can restrict oneself to variations $\bar\delta$
which are tangential to the leaves of the phase space foliation
constructed above. Since $t^a_o$ is a permissible vector field for any
one leaf, we obtain the standard first law
\[\bar\delta M_\Delta = (1/8\pi G)\kappa_{(t_o)}\, \bar\delta a_\Delta
+ \Phi_{(t_o)}\,\bar\delta \tilde Q_\Delta\]
for the restricted variations. The idea of using such restricted
variations was suggested in \cite{cs} in the context of Yang-Mills
fields (although the foliations and other details were not spelled out
there).

To summarize, because there is now a three parameter family of static
solutions rather than two ---or, more precisely, because the standard
surface gravity $\kappa_{(\xi)}$ in static space-times depends on
$a_\Delta, \tilde{Q}_\Delta$ \textit{and} $Q_\Delta$--- a canonical,
permissible evolution field is no longer available. However, there is
still a multitude of permissible evolution fields and corresponding
first laws. Furthermore, one can still define a canonical mass function
$M_\Delta$ on the entire phase space.

\subsection{Yang-Mills fields}
\label{s7.C}

In Einstein-Maxwell theory, with and without the dilaton, there is no
way to construct a quantity with the dimensions of mass from the
fundamental constants in the theory.  The situation is different for
Einstein-Yang-Mills theory because the coupling constant $g$ has
dimensions $(LM)^{-1/2}$.  The existence of a quantity with units of
mass has interesting consequences which we will now discuss.

Let us begin with a summary of the known static solutions in Yang-Mills
theory. First, the Reissner-Nordstr\"om family constitutes a continuous
two parameter set of static solutions of the Einstein-Yang-Mills
theory, labelled by $(a_\Delta, Q^{\YM}_\Delta)$. In addition, there is
a 1-parameter family of `embedded Abelian solutions' with (a fixed)
magnetic charge $P_\Delta^o$, labelled by $(a_\Delta, P_\Delta^o)$.
Finally, there are families of `genuinely non-Abelian solutions'. For
these, the analog of the Israel theorem for Einstein-Maxwell theory
fails to hold \cite{kk}; the theory admits static solutions which need
not be spherically symmetric. In particular, an infinite family of
solutions labelled by two integers $(n_1, n_2)$ is known to exist. All
static, spherically symmetric solutions are known and they correspond
to the infinite sub-family $(n_1, n_2 =0)$, labelled by a single
integer. However, the two parameter family is obtained using a specific
ans\"atz, so there may well exist other static solutions. Although the
available information on static solutions is quite rich, in contrast to
the Einstein-Maxwell-Dilaton system, one is still rather far from
having complete control of the static sector of the Einstein-Yang-Mills
theory.

The zeroth and first laws do hold in the Einstein-Yang-Mills case.  At
present, however, we can only hope to repeat the strategy used in the
last two sub-sections to define a canonical mass function $M_\Delta$ on
portions of the phase space. In order to define it on the full phase
space, of the `uniqueness' and `completeness' conjectures of \cite{cs}
will have to hold (possibly with a suitable modification.%
\footnote{For example it may be appropriate to restrict oneself to the
class of space-times admitting isolated horizons which are complete in
the future. Physically, this is the most interesting case since such
horizons would represent the asymptotic geometry resulting from a
gravitational collapse or black hole mergers.}
Nevertheless, new insight into the static solutions can be obtained by
restricting attention to certain leaves of the phase space.  The basic
idea is taken from \cite{cs} but applied in a slightly different manner
to the more general context of distorted horizons.

Consider a connected component of the known static solutions, labelled
by $\n \equiv (n_1, n_2)$. This is a 1-dimensional sub-space of the
phase space which we denote $S_\n$. Each point in $S_\n$ can be
labelled by the value of the horizon area $a_\Delta$. Calculate the
surface gravity $\kappa_{(\xi)}$ for this family, where $\xi^a$ is the
static Killing field which it unit at infinity and set $\kappa_o =
\kappa_{(\xi)}$ in the construction sketched in Section \ref{s7.A}. We
then obtain a live vector field $t^a_o$, and the corresponding first
law
\[ \delta E^{t_o}_\Delta = \frac{1}{8\pi G}\, \kappa_{(t_o)}\delta
a_\Delta\]
on the full phase space.

When restricted to $S_\n$, we can interpret $E^{t_o}_\Delta$ as the
horizon mass $M_\Delta^{(\n)}$ and replace $\kappa_{(t_o)}$ by the
function $\beta_{\n}(a_{\Delta})$ used in the literature: $\beta_{\n} =
2 \kappa_{(t_o)}R_\Delta$. Then, by integrating the first law along
$S_\n$, one obtains:
\[ M_\Delta^{(\n)} = \frac{1}{2G}\, \int_0^{R_\Delta} \beta_{\n}(x) dx\]
where, we have used the fact that, since $E^{(t_o)}_\Delta$ is a
surface integral at $\Delta$, it vanishes as the horizon area goes to
zero. Thus, the horizon mass is completely determined by
$\beta_\n(a_{\Delta})$.

Next, we use the fact that the Hamiltonian given by $H_{t_o} =
E^{t_o}_{\rm ADM} - E^{t_o}_{\rm \Delta}$ (see (\ref{ham})) is constant
on each connected, static sector if $t_o^a$ coincides with the static
Killing field on the entire sector. By construction, our $t_o^a$ has
this property. In the Einstein-Maxwell case, since there is no constant
with the dimension of energy, it follows that the restriction of
$H^{t_o}_{\rm ADM}$ to the static sector must vanish. The situation is
quite different in Einstein-Yang-Mills theory where the Yang-Mills
coupling constant $g$ provides a scale. In  $c = 1$ units,
$(g\sqrt{G})^{-1} \sim \rm Mass$. Therefore, we can only conclude
\[ M_{\rm ADM}^{(\n)} = M_{\Delta}^{(\n)} + (g\sqrt{G})^{-1} C^{(\n)} \]
for some $\n$-dependent constant $C^{(\n)}$. As the horizon radius
shrinks to zero, the static solution \cite{kk,ym} under consideration
tends to the solitonic solution with the same `quantum numbers' $\n$.
Hence, by taking this limit, we conclude $(g\sqrt{G})^{-1} C^{(\n)} =
M_{\rm ADM}^{\mathrm{soliton},(\n)}$. Therefore, we have the following
interesting relation between the black hole and solitonic solutions:
\[ M_{\rm ADM}^{{\rm BH}, (\n)} = \frac{1}{2G}\, \int_0^{R_\Delta}
\beta_{\n}(x) dx + M_{\rm ADM}^{{\rm soliton}, (\n)} \]
where the integral of $\beta_{\n}$ is evaluated on the 1-dimensional
`parameter space' of $S_{\n}$ (given by the horizon radius).
Furthermore, as is clear from the above discussion, the ADM mass of the
soliton is a multiple of $(g\sqrt{G})^{-1}$. Thus, somewhat
surprisingly, the derivation of the first law in the isolated horizon
framework has led to an interesting relation between the ADM masses of
black holes and their solitonic analogs in the \textit{static sector}.

\section{Discussion}
\label{s8}

In the first part of this paper, we introduced the notions of weakly
isolated and isolated horizons. In contrast with earlier work
\cite{abf2,ac,cs}, the definitions allow for the possible presence of
distortion and rotation at the horizon. In addition, the present
definitions are more geometric and intrinsic; in particular, they never
refer to a foliation.

The notion of an isolated horizon, unlike that of an event horizon, is
completely quasi-local. One can test if a given 3-surface in space-time
is (weakly) isolated or not simply by examining space-time geometry at
the surface. Furthermore, space-times admitting an isolated horizon
$\Delta$ need not admit \textit{any} Killing field even in a
neighborhood of $\Delta$. In particular, they can admit radiation in
the exterior region. Therefore, such space-times can serve as more
realistic models of late stages of a gravitational collapse or black
hole merger. In \cite{abl1} the near $\Delta$ geometry of vacuum
solutions is examined in detail using similar techniques to those used
at null infinity. The resulting structure --- presence of a preferred
`rest frame', constrains on possible isometries, Bondi-type expansions
of the metric --- should be useful to extract physics in the strong
field regime of general relativity, especially in the problem of binary
black hole collisions.

This paper, however, focused on another aspect of isolated horizons:
extensions of the zeroth and first laws of black hole mechanics. All
previous discussions of these laws were restricted to perturbations of
stationary black holes. Using Lagrangian and Hamiltonian frameworks, we
extended these laws to arbitrary space-times admitting non-rotating
isolated horizons in Einstein-Maxwell-dilaton and Einstein-Yang-Mills
theory. Furthermore, the analysis suggests that it should be rather
easy to incorporate other forms of matter, provided they admit
Lagrangian and Hamiltonian descriptions.

The generalization of black hole mechanics presented in this paper has
several interesting features. First, all quantities that enter the
first laws are defined \textit{locally} at the horizon $\Delta$. In
standard treatments, some quantities such as area and surface gravity
are defined at the horizon.  Others, like energy and sometimes
\cite{sw} even the Yang-Mills/Maxwell charge and potential, are
evaluated at infinity. In part because of this `mismatch', to our
knowledge the `physical process version' of the first law \cite{w2} had
not previously been established for processes which change the charge
of the black hole. Since all quantities in the present treatment are
defined locally at the horizon, it is now straightforward to establish
the law for such processes \cite{abf2}. Secondly, other treatments
based on a Hamiltonian framework \cite{w2,w1,sw} often critically use
the bifurcate 2-surface which does not exist in the extremal case.
Therefore, extremal black holes are often excluded from the first law.
The present analysis never makes reference to bifurcate surfaces (which
do not exist in physical space-times resulting from gravitational
collapse). Therefore, our discussion of the first law holds also in the
extremal case. Thirdly, with obvious modifications of boundary
conditions at infinity, our analysis includes cosmological horizons
where thermodynamic considerations are also applicable \cite{gh}.

Finally, and perhaps most importantly, our analysis sheds new light on
the `origin' of the first law: it arose as a necessary and sufficient
condition for the existence of a Hamiltonian generating time evolution.
A new feature of our framework is the existence of an infinite family
of first laws corresponding to the infinite family of `permissible'
vector fields $t^a$. (A vector field $t^a$ is permissible if it is
Hamiltonian, that is, induces canonical transformations on the phase
space.) In theories where we have sufficient control on the space of
static solutions, such as Einstein-Maxwell, one can select a natural
evolution field $t_o^a$. Corresponding to evolution along this $t_o^a$,
there is a canonical notion of energy which can be interpreted as the
mass of the isolated horizon.  There exist also preferred values of
surface gravity and electric potential and a canonical first law. This
additional structure is extremely useful in other applications of the
framework, such as extraction of physical information from numerical
simulations of black-hole collisions. However, it is not essential to
the discussion of mechanics: our derivation of the first law in
Sections \ref{s5} and \ref{s6} does \textit{not} require any knowledge
of the static sector of the theory.

The Hamiltonian approach to black hole mechanics has appeared in the
literature before, most notably in the work of Brown and York
\cite{by}.  The spirit of the Brown-York approach is similar to ours.
In particular, they do not restrict themselves to stationary
situations.  However, in that work, the focus is on an outer,
time-like boundary whereas our focus is on the inner, null boundary
representing the isolated horizon.  Conserved quantities in presence
of internal boundaries were recently discussed also by Julia and Silva
\cite{js} in a more general context of theories with gauge symmetries.
As in our framework, their treatment exploits the simplifications that
occur in a first order formalism and the final surface-integral
expressions of conserved charges are dictated by the precise boundary
conditions imposed at the internal boundaries.  Their treatment is
based on superpotentials and is thus complements the Hamiltonian
methods used here and in \cite{by}.

In this paper, the Lagrangian and Hamiltonian frameworks are based on
real tetrads and Lorentz connections. It is therefore quite
straightforward to extend our analysis to any space-time dimension.
Indeed, it has already been extended to 2+1 dimensions in \cite{adw}.
However, our phase space --- and especially the explicit symplectic
structure used here --- is tailored to the Einstein-matter system.
While it should be possible to extend it to higher derivative theories
of gravity as in \cite{w1}, that task would not be as simple.

\section*{Acknowledgments}

We are grateful to Chris Beetle and Jerzy Lewandowski for countless
discussions. We thank Alejandro Corichi and Daniel Sudarsky for
stimulating correspondence, Jiri Bicak and Werner Israel for pointing
out references on distorted black holes and Thibault Damour, Sean
Hayward and Bob Wald for helpful comments. This work was supported in
part by the NSF grants PHY94-07194, PHY95-14240, INT97-22514 and by the
Eberly research funds of Penn State. SF was supported in part by a
Braddock Fellowship.

\begin{appendix}

\section{Examples of distorted horizons}
\label{a1}

Because of the no hair theorems in the Einstein-Maxwell theory,
distorted horizons have received a rather limited attention in the
literature.  Therefore, in this appendix we will discuss a few explicit
examples in Einstein-Maxwell theory.  For a general construction and an
existence result, see \cite{abl1,jl}.

To obtain explicit solutions, one has to impose symmetries.  All
solutions considered in this section will be static and axi-symmetric.
As one would expect from the no-hair theorems, they fail to be
asymptotically flat, whence they fail to represent isolated black holes
in the standard sense. Nonetheless, they all satisfy the isolated
horizon boundary conditions.  That framework also serves to `explain'
the otherwise surprising feature that the surface gravity of these
solutions depends only on the area and the charge and is insensitive to
the distortion parameters.

In the literature on static, distorted black holes, it is generally
assumed that the solution is valid only in a finite region around the
horizon and its distant behavior is suitably modified by the far-away
matter which causes the distortion.  For undistorted isolated horizons,
Robinson-Trautman space-times \cite{c} offer interesting examples of
vacuum, asymptotically flat solutions which admit isolated horizons but
no Killing fields whatsoever.  Distorted analogs of these solutions are
not known but presumably exist.  It would be interesting to find them.

\subsection{Black Hole in a magnetic universe}

Let us begin with a simple example: an uncharged black hole in an
`external magnetic field' which distorts the horizon.  The specific
solution we wish to consider is static and axisymmetric and was first
obtained in the Ernst-potential framework \cite{fe}. The magnetic field
is non-zero on the horizon. Thus, one has to consider the full set of
Einstein-Maxwell equations on the horizon.

The space-time has topology $S^2\times \Rbar^2$ and the metric is given
by
\begin{equation}\label{melvin}
ds^{2} = F^{2}\left[-\left(1-\frac{2M}{r}\right)dt^{2} + \frac{dr^{2}}{1-2M/r}
+ r^{2}d\theta^{2}\right] + \frac{r^{2}\sin ^{2}\theta }{F^{2}} d\phi^{2}
\end{equation}
where
\[ F = 1+\frac{1}{4}B_{0}^{2}r^{2}\sin ^{2}\theta . \]
$B_0$ is a constant and on the axis the magnetic field is given by $B =
B_0 dr$. Because $F$ diverges at infinity, the metric fails to be
asymptotically flat. For $M=0$, the metric reduces to that of the
Melvin universe and for $M\not= 0$ it admits a Killing horizon at $r =
2M$.  To examine the behavior at the horizon, let us first cast the
metric in the Eddington-Finkelstein coordinates $(v,r,\theta,\phi)$
where $dv = dt + (1- \textstyle{2M \over r})^{-1}\,dr$:
\begin{equation}\label{efmelvin}
ds^{2} = -F^{2}\left(1-\frac{2M}{r}\right)dv^{2} + 2F^{2}dvdr +
F^{2}r^{2}d\theta^{2} + \frac{r^{2}\sin ^{2}\theta}{F^{2}}d\phi^{2}.
\end{equation}
Since the metric is not asymptotically flat, the standard procedure of
normalizing the Killing field to be unit at infinity is not applicable.
Thus, we only have an equivalence class $[\l]$ of (preferred) null
normals to the horizon, $\ell \propto \frac{\partial}{\partial v}$. Let
$\Delta$ be the Killing horizon and assume the Killing field
$\frac{\partial}{\partial v}$ is a member of the equivalence class
$[\ell]$.  It follows trivially that $(\Delta, [\l])$ is a non-rotating
isolated horizon. If $B_0 \not= 0$, the scalar curvature ${}^2\!R$ of
the horizon 2-metric has $\theta$-dependence; the horizon is distorted.
However, an explicit calculation shows that, as in the Schwarzschild
space-time, the surface gravity $\kappa$ is given by $1/2r \equiv
1/2R_\Delta$ and the electrostatic potential $\Phi$ vanishes on
$\Delta$.  At first, it is quite surprising that while the presence of
distortion affects $\mu$, $\Phi_{11}$, $\Psi_2$ and ${}^2\!R$, it does
not affect $\kappa$ or $\Phi$. However, as we saw in Section \ref{s5},
this result is to be expected from the general Hamiltonian
considerations.

\subsection{Distorted black holes as special cases of
Weyl solutions}

In this sub-section, we will review the construction of a large family
of distorted black holes starting from Weyl solutions \cite{distref}
and a recent generalization  of these results to include electric
charge \cite{fk}.

A general class of static, axisymmetric spacetimes was found by Weyl in
1917 \cite{hw}.  The metric for such a spacetime can be cast in the
following form:
\begin{equation} \label{weylmetric}
ds^{2}=-e^{2\psi}dt^{2} + e^{2(\gamma-\psi)}(d\rho^{2}+dz^{2}) +
e^{-2\psi} \rho^2 d\phi^{2}
\end{equation}
where $\psi$ and $\gamma$ are smooth functions of $\rho$ and $z$.
Einstein's vacuum equations expressed in terms of $\psi$ and $\gamma$
take a particularly simple form.  The equation for $\psi$,
\begin{equation} \label{psi}
\psi_{,\rho\rho} + \frac{\psi_{,\rho}}{\rho} + \psi_{,zz} = 0,
\end{equation}
is simply the Laplace equation in \textit{flat space} with cylindrical
coordinates $(\rho, z, \phi)$. (In addition, $\psi$ has to be
independent of the angular coordinate $\phi$.)  Given a solution for
$\psi$, the function $\gamma$ can be determined by simple integration:
\begin{eqnarray}\label{gamma}
\gamma_{,\rho} &=& \rho[ \psi_{,\rho}^{2} - \psi_{,z}^{2}] \nonumber\\
\gamma_{,z} &=& 2\rho[\psi_{,\rho}\psi_{,z}]
\end{eqnarray}

The Schwarzschild metric is of course a particular solution to these
equations and corresponds to choosing for $\psi$ and $\gamma$:
\begin{equation}
\psi = \psi_{\rm S} := \frac{1}{2}\ln\, \left(\frac{L-M}{L+M}\right)
\qquad \mbox{and} \qquad
\gamma = \gamma_{\rm S} := \frac{1}{2}\ln\,
\left(\frac{L^{2} - M^{2}}{L^{2}-\eta^{2}}\right)\, ,
\end{equation}
where $L = \textstyle{1\over 2} (l_{+} + l_{-}), \eta =
\textstyle{1\over 2} (l_{+} - l_{-})$ with $l_{+} = \sqrt{\rho^{2} +
(z+M)^{2}}$ and $l_{-} = \sqrt{\rho^{2} + (z-M)^{2}}$ and $M$ is the
mass of the Schwarzschild solution.  Note that $\psi$ is just the
Newtonian potential due to a rod of length $2M$ placed symmetrically
about the origin on the z-axis. Both $\psi_S$ and $\gamma_S$ diverge
logarithmically in the limit $\rho \rightarrow 0$ (for $\mid z \mid
\leq M$). In order to recast this solution in the standard
Schwarzschild form, one must transform from $(z, \rho)$ to the
Schwarzschild coordinates $(r, \theta)$ by
\begin{equation}\label{schcoord}
z = (r-M)\cos \theta \qquad \mbox{and} \qquad \rho^{2} =
r^{2}(1-2M/r)\sin ^{2}\theta
\end{equation}
This coordinate transformation shows that the horizon, $r=2M$,
corresponds to the line segment $\rho =0$, $\mid z \mid \leq M$ in Weyl
coordinates.  Therefore, the Weyl coordinates cover only the exterior
of the horizon.

Now, the key point is that (\ref{psi}), the only field equation one has
to solve, is linear.  Hence we can `distort' the Schwarzschild solution
simply by adding to $\psi_{\rm S}$ any solution $\psi_{D}$ of the flat
space Laplace equation which is regular along the $z-$axis
\cite{distref}. Thus, we can set
\begin{equation}\label{dist}
\psi = \psi_{\rm S} + \psi_{D} \qquad \mbox{and} \qquad \gamma
=\gamma_{\rm S} + \gamma_{D}.
\end{equation}
Substituting these expressions into (\ref{psi}) and (\ref{gamma}) and
using the forms of the Schwarzschild functions $\psi_S$ and $\gamma_S$,
one can show that at $\rho = 0$,
\begin{equation}\label{gamma2psi}
 \gamma_{D}\mid_{\rho=0} \= 2\psi_{D}\mid_{\rho=0}.
\end{equation}
This fact plays an important role in analyzing the horizon structure.

In Schwarzschild coordinates, the distorted metric takes the form
\begin{equation}\label{metric}
ds^{2} = -e^{2\psi_{D}}(1-2M/r)dt^{2}
+\frac{e^{2(\gamma_{D}-\psi_{D})}}{(1-2M/r)}dr^{2}
+e^{2(\gamma_{D}-\psi_{D})}r^{2}d\theta^{2} + r^{2}\sin ^{2}\theta
e^{-2\psi_{D}}d\phi^{2}.
\end{equation}
As usual, the metric has a coordinate singularity at $r=2M$.  Let us
therefore introduce the Eddington-Finkelstein coordinate $v$ as before.
The metric can be re-expressed in $(v,r,\theta,\phi)$ coordinates as
\begin{eqnarray}\label{efmetric}
ds^{2} &= -e^{2\psi_{D}}(1-2M/r)dv^{2} + (1-2M/r)^{-1}
e^{2\psi_{D}}(e^{2(\gamma_{D}-2\psi_{D})}-1)dr^{2} \nonumber
\\ & + 2e^{2\psi_{D}}dv dr + e^{2(\gamma_{D}-\psi_{D})}r^{2}d\theta^{2}
+ e^{-2\psi_{D}}r^{2}\sin ^{2}\theta d\phi^{2}.
\end{eqnarray}
Using condition (\ref{gamma2psi}) it is not difficult to show that the
coefficient of $dr^2$ in the metric is regular at $r=2M$
\cite{distref}.

It is immediately obvious from (\ref{efmetric}) that the $r=2M$ surface
is a Killing horizon of $\frac{\partial}{\partial v}$. However, we can
not select a preferred normalization for this vector field since the
metric is not asymptotically flat. As in the last sub-section, let
$\Delta$ be the Killing horizon and choose $\ell \propto
\frac{\partial}{\partial v}$.  Then, it is straightforward to verify
that $(\Delta, [\l])$ is a complete, non-rotating isolated horizon. Let
us calculate the value of surface gravity for $\ell \hateq
\frac{\partial}{\partial v}$. We obtain
\begin{equation}
   \kappa \= (e^{2\psi_{D}-\gamma_{D}})\frac{1}{2r} \= \frac{1}{2r}
\end{equation}
where we arrived at the last expression by using (\ref{gamma2psi}).
Again, while spin coefficient $\RePt\mu$, the Weyl component $\Psi_2$
and the scalar curvature ${}^2\!R$ of the horizon metric all depend on
the distortion function $\psi_D$, somewhat surprisingly the surface
gravity $\kappa_\ls$ does not.

The natural question is whether the above framework can be extended to
obtain distorted Reissner-Nordstr\"om solutions.  This turns out to be
non-trivial because the key equation (\ref{psi}) now acquires a source
term from the electromagnetic field and this field itself depends
non-trivially on $\psi$ through the Maxwell equations. At first, the
coupled system appears to be hopelessly difficult.  However, there
exists a prescription \cite{gha} for defining a new potential
$\tilde{\psi}$ in terms of $\psi$ and the electromagnetic field such
that $\tilde{\psi}$ satisfies the flat space Laplacian (\ref{psi}).
Using this method, the known distorted black hole solutions were
recently generalized to the charged case \cite{fk}. The distorted
Reissner-Nordstr\"om solution is given by the metric
\begin{eqnarray}\label{rnmetric}
 ds^{2} &=& -(1- 2M/r + Q^{2}/r^{2})e^{2\psi_{D}}dt^{2} +
{(1- 2M/r + Q^{2}/r^{2})^{-1}}e^{2(\gamma_{D} - \psi_{D})}dr^{2}
\nonumber\\
&& + e^{2(\gamma_{D} - \psi_{D})}r^{2}d\theta^{2} +
e^{-2\psi_{D}}r^{2}\sin ^{2}\theta d\phi^2 \, .
\end{eqnarray}
The forms of $\psi_D$ and $\gamma_D$ are now substantially more
complicated than in the uncharged case. Nonetheless, it is still
possible to show that (\ref{gamma2psi}) continues to hold. As before
this equality implies that the apparent singularity at $r_H^2-2Mr_H+Q^2
= 0$ is only a coordinate singularity.  The surface defined by $r=r_H$
is a Killing horizon of $\frac{\partial}{\partial t}$. There is once
again, no natural way to normalize the Killing field, so we only have
an equivalence class $[\l^a]$ of null normals to the Killing horizon.
$(\Delta, [\l])$ is a non-rotating isolated horizon.

The surface gravity of $\frac{\partial}{\partial t}$ is given by
\begin{equation}\label{drnsurfgrav}
\kappa = \frac{1}{2r_H}\left(1- \frac{Q^2}{r_H^2}\right).
\end{equation}
Again, the surface gravity is independent of the distortion of the
horizon and has the same dependence on the horizon radius $R_{\Delta}$
(which turns out to be equal to $r_H$) and charge $Q$ as in
Reissner-Nordstr\"om spacetime.  Considerations of Section \ref{s5}
suggest this peculiar behavior of $\kappa$ in all these examples is
not accidental but can be `explained' from general Hamiltonian
considerations which led us to the first law.

\section{The Newman-Penrose formalism}
\label{appb}
\subsection{Notation and Conventions}

Let us begin with a summary of the Newman-Penrose formalism (see
\cite{np} or \cite{pr,sc,s} for a complete account).  Apart from the
spacetime signature which we take to be $(-,+,+,+)$, we will follow the
conventions used in \cite{s}. Consider a tetrad of null vectors $n$,
$\ell$, $m$ and $\bar{m}$ ($n$ and $\ell$ are real while $m$ is
complex) which satisfy
\begin{equation}
\begin{array}{rcl@{\hspace{5mm}}rcl@{\hspace{5mm}}rcl}
n.\ell &=& -1 & n.m &=& 0 & n.\bar{m} &=& 0 \\ \ell .m &=& 0 & \ell
.\bar{m} &=& 0 & m.\bar{m} &=& 1 \, .
\end{array}
\end{equation}
The directional derivatives along the basis vectors are denoted by
\begin{equation}
    D = \ell^{a} \nabla_{a} \qquad \qquad
    \Delta = n^{a} \nabla_{a}  \qquad \qquad
    \delta = m^{a} \nabla_{a}  \qquad \qquad
    \bar{\delta} = \bar{m}^{a} \nabla_{a} \, .
\end{equation}
The full the information contained in the connection is expressed in
terms of twelve complex scalars called the Newman-Penrose spin
coefficients defined as follows:

\begin{equation}
\begin{array}{rcl@{\hspace{5mm}}rcl@{\hspace{5mm}}rcl}
     \kappa &=& -m^{a}\ell^{b}\nabla_{b}\ell_{a} &
     \epsilon &=& \frac{1}{2}(\bar{m}^{a} \ell^{b}\nabla_{b}m_{a}
            - n^{a}\ell^{b}\nabla_{b}\ell_{a}) &
     \pi &=& \bar{m}^{a}\ell^{b}\nabla_{b}n_{a}\\[1.5mm]
     \sigma &=& -m^{a}m^{b}\nabla_{b}\ell_{a}  &
     \beta &=& \frac{1}{2}(\bar{m}^{a}m^{b}\nabla_{b}m_{a}
            - n^{a}m^{b}\nabla_{b}\ell_{a})  &
     \mu &=& \bar{m}^{a}m^{b}\nabla_{b}n_{a}\\[1.5mm]
     \rho &=& -m^{a}\bar{m}^{b}\nabla_{b}\ell_{a}   &
     \alpha &=& \frac{1}{2}(\bar{m}^{a}\bar{m}^{b}\nabla_{b}m_{a}
            - n^{a}\bar{m}^{b}\nabla_{b}\ell_{a}) &
     \lambda &=& \bar{m}^{a}\bar{m}^{b}\nabla_{b}n_{a}\\[1.5mm]
     \tau &=& -m^{a}n^{b}\nabla_{b}\ell_{a} &
     \gamma &=& \frac{1}{2}(\bar{m}^{a}n^{b}\nabla_{b}m_{a}
            -n^{a}n^{b}\nabla_{b}\ell_{a})  &
     \nu &=& \bar{m}^{a}n^{b}\nabla_{b}n_{a} \,  .
\end{array}
\end{equation}
It is sometimes more useful to express these definitions in terms of
covariant derivatives of the basis vectors:
\begin{equation}
\begin{array}{rcl@{\hspace{5mm}}rcl}
    D\l &=& (\epsilon + \bar{\epsilon})\l - \bar{\kappa}m - \kappa \bar{m} &
    Dn &=& -(\epsilon + \bar{\epsilon})n + \pi m + \bar{\pi} \bar{m}\\[1.5mm]
    \Delta \l &=& (\gamma + \bar{\gamma})\l - \bar{\tau}m - \tau \bar{m} &
    \Delta n &=& -(\gamma + \bar{\gamma})n + \nu m + \bar{\nu} \bar{m}\\ [1.5mm]
    \delta \l &=& (\bar{\alpha} + \beta)\l - \bar{\rho}m - \sigma \bar{m} &
    \delta n &=& -(\bar{\alpha} + \beta)n +\mu m + \bar{\lambda} \bar{m}\\ [1.5mm]
    Dm &=& \bar{\pi}\l - \kappa n + (\epsilon - \bar{\epsilon})m &
    \Delta m &=& \bar{\nu}\l - \tau n + (\gamma - \bar{\gamma})m\\[1.5mm]
    \delta m &=& \bar{\lambda}\l - \sigma n + (\beta - \bar{\alpha})m &
    \bar{\delta} m &=& \bar{\mu}\l - \rho n + (\alpha - \bar{\beta})m
    \, .
\end{array}
\end{equation}
The ten independent components of the Weyl tensor are expressed in
terms of five complex scalars $\Psi_{0}$, $\Psi_{1}$, $\Psi_{2}$,
$\Psi_{3}$ and $\Psi_{4}$. The ten components of the Ricci tensor are
defined in terms of four real and three complex scalars $\Phi_{00}$,
$\Phi_{11}$, $\Phi_{22}$, $\Lambda$, $\Phi_{10}$, $\Phi_{20}$ and
$\Phi_{21}$ . These scalars are defined as follows:
\begin{equation}
\begin{array}{rcl@{\hspace{5mm}}rcl@{\hspace{5mm}}rcl}
    \Psi_{0}&=& C_{abcd}\ell^{a}m^{b}\ell^{c}m^{d} &
    \Phi_{01} &=&\frac{1}{2}R_{ab}\ell^{a}m^{b} &
    \Phi_{10} &=&\frac{1}{2}R_{ab}\ell^{a}\bar{m}^{b} \\[1.5mm]
    \Psi_{1} &=&C_{abcd}\ell^{a}m^{b}\ell^{c}n^{d} &
    \Phi_{02} &=&\frac{1}{2}R_{ab}m^{a}m^{b} &
    \Phi_{20} &=&\frac{1}{2}R_{ab}\bar{m}^{a}\bar{m}^{b}  \\[1.5mm]
    \Psi_{2} &=& C_{abcd}\ell^{a}m^{b}\bar{m}^{c}n^{d}  &
    \Phi_{21} &=&\frac{1}{2}R_{ab}\bar{m}^{a}n^{b}  &
    \Phi_{12} &=&\frac{1}{2}R_{ab}m^{a}n^{b} \\[1.5mm]
    \Psi_{3} &=&C_{abcd}\ell^{a}n^{b}\bar{m}^{c}n^{d}  &
    \Phi_{00} &=&\frac{1}{2}R_{ab}\ell^{a}\ell^{b}  &
    \Phi_{11} &=&\frac{1}{4}R_{ab}(\ell^{a}n^{b} + m^{a}\bar{m}^{b}) \\[1.5mm]
    \Psi_{4} &=&C_{abcd}\bar{m}^{a}n^{b}\bar{m}^{c}n^{d}  &
    \Phi_{22} &=&\frac{1}{2}R_{ab}n^{a}n^{b}   &
    \Lambda &=& \frac{R}{24} \, .
\end{array}
\end{equation}
The six components of the Electromagnetic-Field $2-$form $\emF_{ab}$
can be defined in terms of three complex scalars:
\begin{equation}
    \phi_{0} = -\ell^{a}m^{b}\emF_{ab} \qquad
    \phi_{1} = -\half (\ell^{a}n^{b} - m^{a}\bar{m}^{b})\emF_{ab}
    \qquad \phi_{2} =n^{a}\bar{m}^{b}\emF_{ab} \, .
\end{equation}
The eight real Maxwell equations $d\emF = 0$ and $d \dual{\emF} = 0$
can be written as a set of four complex equations:
\begin{eqnarray} \label{maxwelleq}
    D\phi_{1} - \bar{\delta}\phi_{0} &=& (\pi - 2\alpha)\phi_{0}
              +2\rho\phi_{1} - \kappa\phi_{2}  \\
    D\phi_{2} - \bar{\delta}\phi_{1} &=& -\lambda\phi_{0} + 2\pi\phi_{1}
              + (\rho - 2\epsilon)\phi_{2}  \\
    \Delta\phi_{0} - \delta\phi_{1} &=& (2\gamma - \mu)\phi_{0}
              - 2\tau\phi_{1} + \sigma\phi_{2} \\
    \Delta\phi_{1} - \delta\phi_{2} &=& \nu\phi_{0} - 2\mu\phi_{1}
              + (2\beta - \tau)\phi_{2}   \, .
\end{eqnarray}

\subsection{Boundary Conditions}

In this section we describe the isolated horizon boundary conditions in
the Newman-Penrose formalism.  We will restrict ourselves to
Einstein-Maxwell theory with zero cosmological constant.

In a null-tetrad adapted to the null hypersurface $\Delta$, take $\l$
to be a null normal, $m$ and $\bar{m}$ tangent to $\Delta$ and $n$
transverse to $\Delta$.  Since $\l$ is hypersurface orthogonal and
null, it is geodesic. This implies that\footnote{We will denote the NP
spin coefficient $\kappa$ by $\kappa_{NP}$ to distinguish it from the
surface gravity $\kappa_{(\l)}$.} $\kappa_{NP} \= 0$ and $\ImPt{\rho}
\= 0$. Thus
\begin{equation}
D\l^b := \l^{a}\grad_{a}\l^{b} \= (\epsilon + \bar{\epsilon})\l^{b} \,
.
\end{equation}
The surface gravity is therefore given by $\kappa_{(\l)} = \epsilon +
\bar{\epsilon}$ and the expansion of $\l$ is $\theta_{(\l)} \=
\RePt{\rho}$.

For a non-expanding horizon $\Delta$, the conditions on $\l$ imply
$\rho \= 0$ and the Raychaudhuri equation then implies $\sigma \= 0$
and $\Phi_{00} = \half R_{ab}\l^{a}\l^{b} \= 0$.  Furthermore, from
(\ref{pbackf}) (which is a consequence of the energy condition), it
follows that $\phi_{0} \= 0$.  This leads to the following conditions
on the Ricci tensor at the horizon
\begin{equation}
\begin{array}{rcl@{\hspace{10mm}}rcl@{\hspace{10mm}}rcl}
\Phi_{00} &\=& 0 & \Phi_{01} &\=& 0 &\Phi_{10} &\=& 0 \\ \Phi_{02} &\=&
0 & \Phi_{20} &\=& 0 & \Phi_{11} &\=& -2G\phi_{1}\bar{\phi}_{1} \, .
\end{array}
\end{equation}
The first Maxwell equation (\ref{maxwelleq}) gives
\begin{equation}
 D\phi_{1} \= 0 \qquad \rm{which\ implies} \qquad
 D\Phi_{11} \= 0 \, .
\end{equation}
Also, as shown in (\ref{psi0psi1})
\begin{equation}
\Psi_{0} \= 0 \qquad \rm{and} \qquad \Psi_{1} \= 0 \, .
\end{equation}
The intrinsically defined one-form $\omega_{a}$ defined in
(\ref{omegadefn}) is given by
\begin{equation}
\omega_{a} = -\kappa_{(\l)}n_{a} + (\alpha + \bar{\beta})m_{a} +
(\bar{\alpha} + \beta)\bar{m}_{a} \, .
\end{equation}
It is often convenient to choose the null tetrad such that
$\pback{dn}=0$ which implies
\begin{equation}
\mu \= \bar{\mu} \qquad \rm{and} \qquad \pi \= \alpha + \bar{\beta} \,
.
\end{equation}
In this case we get a foliation of $\Delta$ spanned by $m$ and
$\bar{m}$.  Furthermore, by an appropriate spin transformation, we can
choose $\epsilon$ to be real so that $\epsilon \= \bar{\epsilon}$ and
thus the foliation is Lie dragged along $\l$:
\begin{equation}
\Lie_{\l} m^{a} = (\epsilon - \bar{\epsilon})m^{a} \= 0 \, .
\end{equation}
The one-form $\omega$ now becomes
\begin{equation}
\omega_{a} = -\kappa_{(\l)}n_{a} + \pi m_{a} + \bar{\pi}\, \bar{m}_{a}
\, .
\end{equation}
Let us consider a weakly isolated horizon $(\Delta, [\l])$. The
condition $\Lie_{\l}\omega = 0$ is equivalent to requiring
\begin{equation}
\Lie_{\l}\, \pi \= 0 \qquad \rm{and} \qquad \Lie_{\l}\>\kappa_{(\l)} \=
0
\end{equation}
and as we proved in Section \ref{s2.B}, these conditions imply that the
surface gravity $\kappa_{(\l)}$ is constant on $\Delta$.

As mentioned in Section \ref{s2.B}, a weakly isolated horizon with
non-zero surface gravity admits a natural foliation.  In the
Newman-Penrose framework this foliation can be characterized as
follows: It is the unique foliation on each leaf of which the pull-back
of the 1-form $\pi m_a + \bar\pi\, \bar{m}_a $ is divergence-free. This
condition was first introduced by H\'a\'ji\v{c}ek \cite{ph} in the
context of stationary spacetimes.

Finally, since our boundary conditions require that $\Lie_{\l}\,
\kappa_{\ls} \= 0$, in a sense, a part of the zeroth law is simply
assumed.  As mentioned in Section \ref{s2.B}, we could have used a
slightly different set of boundary conditions which make no direct
requirement on $\kappa_{\ls}$ and yet lead to the zeroth law (as well
as the results of Sections \ref{s3} - \ref{s7}).

Let $(\Delta,[\l])$ be a non-expanding horizon, equipped with an
equivalence class $[\l]$ of null normals to related to each other by
constant positive rescalings.  As above introduce a null tetrad where
$\l$ is an element of $[\l]$, $m$ and $\bar{m}$ are tangent to the
foliation, $n$ is curl free and $\epsilon$ is real.  In place of
Definition 2, let us assume that $\Delta$ admits a foliation by a
family $S_{\Delta}$ of $2-$spheres transverse to $[\l]$ such that the
NP spin coefficients in an associated null-tetrad satisfy:
\begin{equation}
 \Lie_{\l}\,\mu \= 0 \qquad {\rm and}\qquad \Lie_{\l}\,\pi \= 0 \, .
\end{equation}
These conditions now replace the requirement $\Lie_{\l} \omega \= 0$
used in the definition of a weakly isolated horizon.  We can prove the
zeroth law from these conditions as follows.  First, consider the
definition $2\grad_{[a}\grad_{b]}\xi_{c} = {R_{abc}}^{d}\xi_{d}$ of the
Riemann tensor.  In the NP formalism, these are written as a set of
$18$ complex equations known as the `field equations'.  For our
purposes, we need only three of these equations \cite{s}
\begin{eqnarray}
D\alpha - \bar{\delta}\epsilon &=& (\rho + \bar{\epsilon}
-2\epsilon)\alpha + \beta\bar{\sigma} - \bar{\beta}\epsilon -
\kappa\lambda - \bar{\kappa}\gamma + (\epsilon + \rho)\pi + \Phi_{10}
\nonumber \\ D\beta - \delta\epsilon &=& (\alpha + \pi)\sigma +
(\bar{\rho} - \bar{\epsilon})\beta -(\mu + \gamma)\kappa -
(\bar{\alpha} - \bar{\pi})\epsilon + \Psi_{1} \nonumber\\ D\mu
-\delta\pi &=& (\bar{\rho} - \epsilon -\bar{\epsilon})\mu +
\sigma\lambda + (\bar{\pi} - \bar{\alpha} + \beta)\pi - \nu\kappa +
\Psi_{2} + 2\Lambda \nonumber \, .
\end{eqnarray}
Adding the first equation to the complex conjugate of the second
equation and imposing our boundary conditions gives
\begin{equation}\label{sphsymm}
\delta(\epsilon + \bar{\epsilon}) \= \delta\kappa_{(\l)} \= 0
\end{equation}
while the third equation reduces to
\begin{equation}\label{psi2kappamu}
\Psi_{2} \= (\epsilon + \bar{\epsilon})\mu \, .
\end{equation}
Equation (\ref{sphsymm}) tells us that surface gravity is constant on
each leaf of the foliation. It now only remains to show that it is also
constant along $\l$. To show this we turn to the Bianchi identity:
$\grad_{[a}R_{bc]de} = 0$. In the NP formalism, this is written as a
set of nine complex and two real equations. We shall need only two of
these equations \cite{s}
\begin{eqnarray}
    &&D\Psi_{2} - \bar{\delta}\Psi_{1} + \Delta\Phi_{00} -
    \bar{\delta}\Phi_{01} + 2D\Lambda  \nonumber \\
    && \qquad =  -\lambda\Psi_{0} + 2(\pi-\alpha)\Psi_{1}  + 3\rho\Psi_{2}
    -2\kappa\Psi_{3}+\bar{\sigma}\Phi_{02} \nonumber \\
    && \qquad\quad + (2\gamma + 2\bar{\gamma} - \bar{\mu})\Phi_{00}
    - 2(\alpha +\bar{\tau}) \Phi_{01}
    - 2\tau\Phi_{10} + 2\rho\Phi_{11} \nonumber \\
    \medskip \\
    &&D\Phi_{11} - \delta\Phi_{10} + \Delta\Phi_{00} -
    \bar{\delta}\Phi_{01} + 3D\Lambda  \nonumber \\
    && \qquad = (2\gamma + 2\bar{\gamma} - \mu -\bar{\mu})\Phi_{00}
    + (\pi - 2\alpha - 2\bar{\tau})\Phi_{01}
    +(\bar{\pi} - 2\bar{\alpha} - 2\tau)\Phi_{10} \nonumber  \\
    && \qquad\quad + 2(\rho + \bar{\rho})\Phi_{11}
    + \bar{\sigma}\Phi_{02}
    + \sigma \Phi_{20}-   \bar{ \kappa} \Phi_{12}
    - \kappa\Phi_{21} \nonumber \, .
\end{eqnarray}

Subtracting these equations, imposing our boundary conditions and using
$\Lambda = 0$, we get $D\Psi_{2} \= 0$.  Combining this result with
(\ref{psi2kappamu}) gives $D(\epsilon + \bar{\epsilon}) \= 0$.  This
completes the proof of the zeroth law within the alternate definition
of weak isolation.  Most of the results of this paper were first
obtained using that definition.  However, since that notion is tied so
heavily to the presence of a foliation, its intrinsic meaning is
somewhat obscure. Therefore, it was then replaced by Definition 2 used
in the main body of the paper.

\end{appendix}

\end{document}